\documentclass[final,5p,times,sort&compress]{elsarticle}

\usepackage{amssymb}
\usepackage{amsthm}

\usepackage{multirow}
\usepackage{amsmath}
\usepackage{color}
\usepackage{graphicx}

\journal{Physica E}

\begin{document}

\begin{frontmatter}



\title{Magnetic and magnetocaloric properties of the exactly solvable mixed-spin Ising model \\
on a decorated triangular lattice in a magnetic field}


\author[label1]{Lucia G\'alisov\'a}
\author[label2]{Jozef Stre\v{c}ka}

\address[label1]{Department of Applied Mathematics and Informatics,
					Faculty of Mechanical Engineering, Technical University,
					Letn\'a 9, 042 00 Ko\v{s}ice, Slovakia}
\address[label2]{Department of Theoretical Physics and Astrophysics, Institute of Physics,
                    Faculty of Science, P. J. \v{S}af\'arik University,
                    Park Angelinum 9, 040 01 Ko\v{s}ice, Slovakia}

\begin{abstract}
The ground state, zero-temperature magnetization process, critical behaviour and isothermal entropy change of the mixed-spin Ising model on a decorated triangular lattice in a magnetic field are exactly studied after performing the generalized decoration-iteration mapping transformation. It is shown that both the inverse and conventional magnetocaloric effect can be found near the absolute zero temperature. The former phenomenon can be found in a vicinity of the discontinuous phase transitions and their crossing, while the latter one occurs in some paramagnetic phases due to a spin frustration to be present at zero magnetic field. The inverse magnetocaloric effect can also be detected slightly above continuous phase transitions following the power-law dependence $|-\Delta{\cal S}_{iso}^{min}|\propto h^n$, where $n$ depends basically on the ground-state spin ordering.
\end{abstract}


\end{frontmatter}



\section{Introduction}
\label{sec:1}

Exactly solvable mixed-spin Ising models on two-dimen\-sional (2D) lattices belong to attractive issues of the statistical mechanics, because they provide a convincing evidence for many controversial results predicted in the phase transition theory. More specifically, these systems represent useful testing ground for a rigorous theoretical investigation of the spin frustration~\cite{Gon87, Str06b, Str12, Str13, Poh16, Fis60a, Fis60b, Hat68, Mas73, Gia88, Aza88, Lu05, Can06, Gal16}, reentrant phenomenon~\cite{Gon87, Str06b, Str12, Str13, Poh16, Mas73, Gal16, Mat07, Str06a, Jas05b}, compensation behaviour~\cite{Str04, Jas05b, Str12, Mat07, Lac04, Kar15}, change of the usual critical points to the multi-critical ones~\cite{Str06b, Str06a, Lip95, Hin05, Kar15}, as well as striking spontaneous 'quasi one-dimensional' spin order at the absolute zero temperature~\cite{Str07, Can08}. Moreover, quantum effects can also be exactly examined by imposing the transverse magnetic field or the biaxial single-ion anisotropy on some spins of the 2D lattices~\cite{Lac04, Str03, Str04, Jas05a, Jas05b, Eki09}.

On the other hand, a rigorous investigation of the magnetic-field effect on magnetic properties of the 2D Ising models still remains an open topic due to the lack of closed-form exact solution for the partition function at finite magnetic fields. At present, there are known just a few mixed-spin Ising models which allow an exact study of the magnetic-field effect in 2D, namely, the spin-$1/2$ Ising model on a kagom\'e lattice~\cite{Gia88,Aza88,Lu05}, the spin-$1/2$ Fisher super-exchange model on a square lattice~\cite{Fis60a, Fis60b} and its another extensions~\cite{Hat68, Mas73, Can06, Gal16}. All these models involve the action of the longitudinal magnetic field on two-thirds of all the lattice sites. This assumption allows one to obtain exact solutions for the models by using the concept of generalized algebraic transformations~\cite{Fis59,Dom60,Syo72,Str10}. In fact, the generalized decoration-iteration and star-triangle transformations establish a rigorous mapping correspondence between the afore-mentioned models and the spin-$1/2$ Ising lattices with known exact analytical solutions~\cite{Dom60,Ons44, Hou50}, which gives the opportunity to gain a comprehensive picture on the critical behaviour as well as thermodynamics of these systems. Moreover, it has been demonstrated in our recent works~\cite{Gal16} and~\cite{Gal17} that the Fisher super-exchange model and its another variants represent excellent tools for a rigorous theoretical investigation of the magnetocaloric effect (MCE) in a proximity of the continuous (second-order) phase transitions. Thanks to its exact solvability, important MCE quantities such as the isothermal entropy change and the adiabatic temperature change may be straightforwardly calculated.

Besides the academic interest, the 2D mixed-spin Ising models, which do not assume the magnetic-field effect on all particles, are also valuable for elucidation of an unusual magnetic behaviour of several real compounds. This specific requirement can be observed, e.g. in the ferrimagnet SrCr$_8$Ga$_4$O$_{19}$, which consists of kagom\'e slabs with magnetic spins residing just at one-third of all lattice sites~\cite{Obr88}, or various high-$T_c$ antiferromagnetic cuprates including CuO$_2$ planes in their crystal structures, such as  Ba$_2$YCu$_3$O$_{9-\delta}$~\cite{Sie87}, Pr$_{1-x}$LaCe$_x$CuO$_{4+\delta}$~\cite{Cam15}, La$_{2-x}M_x$CuO$_{4-\delta}$ ($M$=Ba, Sr)~\cite{Vas89}.

Taking into account the aforementioned facts, we propose in this paper a novel mixed-spin Ising model on a decorated triangular lattice in a longitudinal magnetic field for which the closed-form exact solution can be derived. The proposed model is somewhat reminiscent of the generalized Fisher super-ex\-change model on a square lattice~\cite{Fis60a, Fis60b, Hat68, Mas73, Can06, Gal16}, but its main novelty lies in definition on a non-bipartite triangular lattice, which provides a possible playground for a new type of spin frustration.
Besides the ground-state analysis, our attention is focused on the study of the zero-temperature magnetization process, critical behaviour and magnetocaloric properties of the system.

The organization of the paper is as follows. Section~\ref{sec:2} contains a detailed
description of the lattice structure and an exact solution of the model. The numerical  results obtained for two representative spin versions of the model, namely the mixed spin-$(1/2,1)$ model and the mixed spin-$(1/2,3/2)$ model, are presented in Section~\ref{sec:3}. Finally, the paper is closed with a summary of the most interesting findings listed in Section~\ref{sec:4}.
\begin{figure}[t!]
\centering
  \includegraphics[width=0.96\columnwidth]{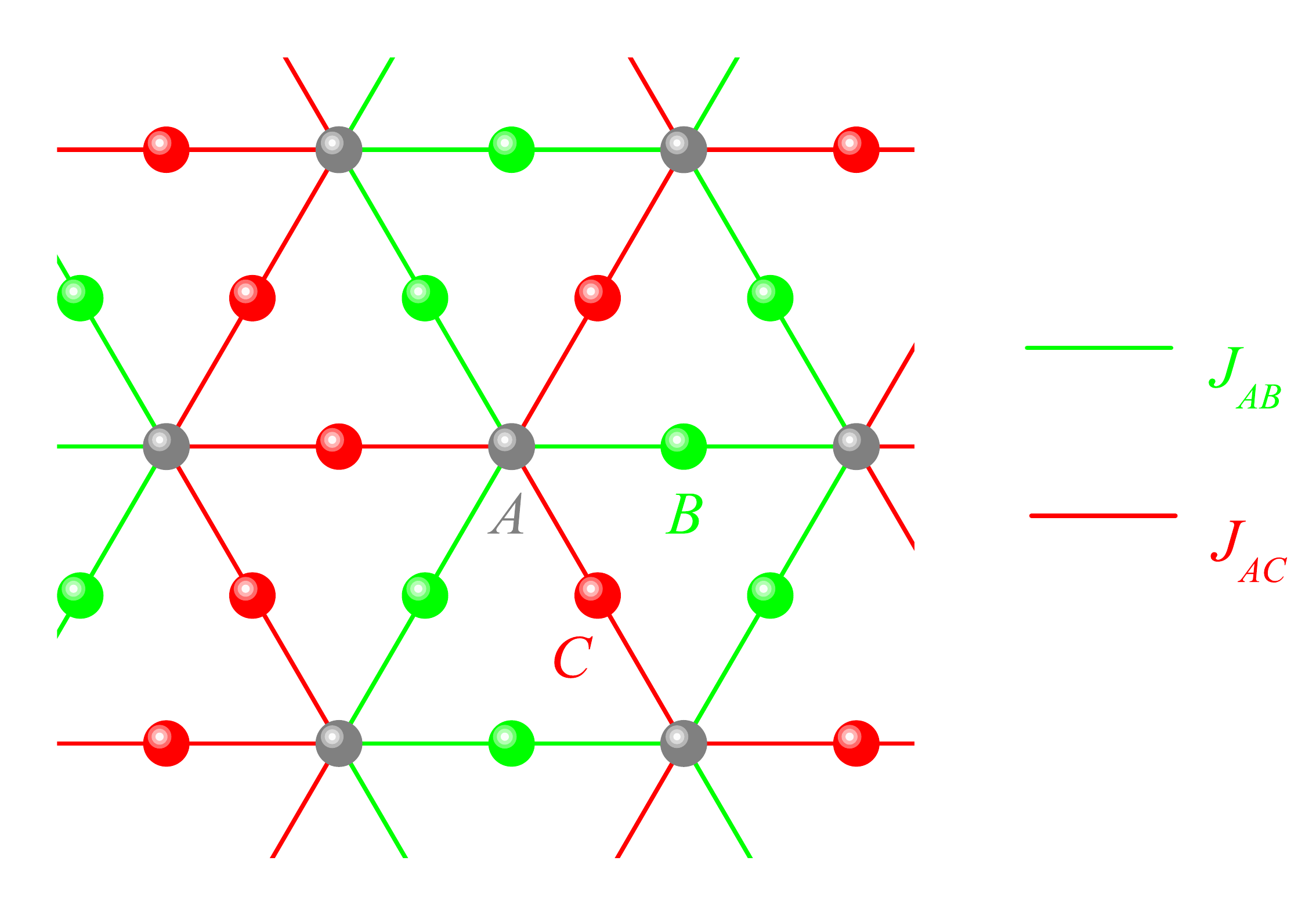}
\caption{(Color online) A part of the mixed-spin Ising model on a decorated triangular lattice. Gray circles denote nodal lattice sites occupied by the spins $\sigma = 1/2$ (sublattice $A$), green and red circles label lattice positions of the decorating spins $s_B\geq1$ and $s_C\geq1$ forming $B$- and $C$-sublattices, respectively.}
\label{fig:1}
\end{figure}

\section{Model and its exact analytical solution}
\label{sec:2}

We consider the spin-$1/2$ Ising model on $N$-site triangular lattice, where a half of the lattice bonds is decorated by the spins $s_{B}\geq1$, while other ones are decorated by the spins $s_{C}\geq1$ in a~manner illustrated in Fig.~\ref{fig:1}. The displayed magnetic structure can be viewed as three-sublattice system, in which the spins of the magnitude $1/2$ localized at nodal lattice sites ($A$-sublattice) are surrounded by three decorating spins of the sublattice $B$ and three decorating spins of the sublattice~$C$. Assuming an uniaxial single-ion anisotropy and a longitudinal magnetic field acting on decorating spins, the total Hamiltonian of the model reads:
\begin{eqnarray}
\label{eq:H_tot}
H\!\!\!&=&\!\!\! -\,J_{AB}\!\!\!{\sum_{i\in A,m\in B}}\!\!\sigma_{i}^{z}s_{m}^{z} - J_{AC}\!\!\!{\sum_{i\in A,n\in C }}\!\!\sigma_{i}^{z}s_{n}^{z}
- D_{B}\!\sum_{m\in B}\left(s_{m}^{z}\right)^{2}
\nonumber\\
\!\!\!& &\!\!\!-\, D_{C}\!\sum_{n\in C}\left(s_{n}^{z}\right)^{2}
- h_B\!\sum_{m \in B}s_{m}^{z}
- h_C\!\sum_{n \in C}s_{n}^{z}.
\end{eqnarray}
In above, $\sigma_{i}^{z}=\pm1/2$, $s_{m}^{z} = -s_{B}, -s_{B}+1,\ldots,s_{B}$ and $s_{n}^{z} = -s_{C}, -s_{C}+1,\ldots,s_{C}$ are the Ising spins from $A$-, $B$- and $C$-sublattices, respectively, the parameters $J_{AB}$ and $J_{AC}$ represent the exchange couplings between the nearest-neighbouring spins from the $A$-, $B$-sublattices and $A$-, $C$-sublattices, respectively, and $D_{B}$, $D_{C}$ stand for the uniaxial single-ion anisotropy acting on the spins from the $B$- and $C$-sublattices. Last two terms $h_B = g_B\mu_{\rm B}h$ and $h_C = g_C\mu_{\rm B}h$ in Eq.~(\ref{eq:H_tot}) label Zeeman energies of the decorating spins $s_{B}$ and $s_{C}$, respectively, which may be in general different due to distinct Land\'e g-factors $g_B$, $g_C$ ($\mu_{\rm B}$ is Bohr magneton and $h$ is the external magnetic field).
For the sake of simplicity, the exact solution for the proposed mixed-spin Ising model will be obtained in the thermodynamic limit $N \to \infty$ by imposing periodic boundary conditions.

In view of further calculations, it is useful to divide the total Hamiltonian~(\ref{eq:H_tot}) into two parts $H= \sum_{m=1}^{3N/2}H_m^{B}  +  \sum_{n=1}^{3N/2}H_n^{\,C}$, where the former (latter) summation runs over the lattice bonds that involve all the interaction terms associated with the decorating spin $s_{B}$ ($s_{C}$) at $m$th ($n$th) site of the sublattice $B$ ($C$):
\begin{equation}
\label{eq:H_k}
H_k^{X}=  -J_{AX}s_{k}^{z}\big(\sigma_{i}^{z} + \sigma_{j}^{z}\,\big) - D_{X}\,\big(s_{k}^{z}\big)^{2}
- h_X s_{k}^{z}\quad (X=B\,\,{\rm or}\,\,C)\,.
\end{equation}
The partition function of the mixed-spin Ising model defined through the Hamiltonian~(\ref{eq:H_tot}) can be partially factorized by performing the summation over spin degrees of freedom of the spins $s_{B}$, $s_{C}$ before summing over all possible spin states of particles in the $A$-sublattice. Moreover, spin-state summations corresponding to particles of the $B$- and $C$-sublattices can be performed independently of each other, because there is no direct interaction between them. As a result, the partition function of the model reads:
\begin{equation}
\label{eq:Z}
Z= \sum_{\left\{\sigma = \pm1/2\right\}}\prod_{m,n}^{3N/2}\sum_{s_{m}^{z}= -s_{B}}^{s_{B}}\!\!\exp{\left(-\beta H_m^{B}\right)}\sum_{s_{n}^{z}= -s_{C}}^{s_{C}}\!\!\exp{\left(-\beta H_n^{\,C}\right)}\,,
\end{equation}
where $\beta = 1/(k_{\rm B}T)$ stands for the inverse temperature ($k_{\rm B}$ is Boltzmann constant and $T$ is the absolute temperature). The structure of Eq.~(\ref{eq:Z}) gives the opportunity to perform the generalized decoration-iteration mapping transformation~\cite{Fis59,Dom60,Syo72,Str10}:
\begin{equation}
\label{eq:DIT}
\sum_{s_{k}^{z}= -s^{X}}^{s^{X}}\!\!\exp{\left(-\beta H_k^{X}\right)}= P_X \exp{\bigg[\beta J_{X}^{\,e\!f\!f}\sigma_{i}^{z}\sigma_{j}^{z} + \frac{\beta h_{X}^{\,e\!f\!f}}{6}\,\big(\sigma_{i}^{z} + \sigma_{j}^{z}\,\big)\bigg]}.
\end{equation}
The idea of this algebraic technique is to replace all interaction parameters associated with the decorating spin $s_k^{X}$ by a new effective interaction $J_{X}^{\,e\!f\!f}$ between the remaining nodal spins $\sigma_{i}^{z}$, $\sigma_{j}^{z}$ and by a new effective field $h_{X}^{\,e\!f\!f}$ acting on these spins. It is worthy to note that the relation~(\ref{eq:DIT}) has to be valid for all spin state combinations of $\sigma_{i}^{z}$ and $\sigma_{j}^{z}$. This 'self-consistency' condition unambiguously determines the set of not yet specified transformation parameters:
\begin{eqnarray}
\label{eq:AJH}
P_X \!\!\!&=&\!\!\! \sqrt[4]{W_X(1)W_X(-1)W_X^{2}(0)}\,,
\nonumber\\
J_{X}^{\,e\!f\!f} \!\!\!&=&\!\!\! k_{\rm B}T\left[\ln W_X(1) + \ln W_X(-1) - 2\ln W_X(0)\right],
\nonumber\\
h_{X}^{\,e\!f\!f} \!\!\!&=&\!\!\! 3k_{\rm B}T\left[\ln W_X(1) - \ln W_X(-1)\right].
\end{eqnarray}
The values of the functions $W_X(\pm1)$ and $W_X(0)$ depend on the magnitude of decorating spin, model parameters and the temperature according to the formula:
\begin{equation}
\label{eq:W}
W_X(a)  =\!\!\! \sum_{s_{k}^{z}= -s_{X}}^{s_{X}}\!\!\exp{\left[\beta D_X \big(s_{k}^{z}\big)^{2}\right]}\cosh\left[\beta s_{k}^{z}(a J_{AX} -h_X)\right].
\end{equation}
After a substituting Eq.~(\ref{eq:DIT}) into~(3), one gains a rigorous relation
\begin{eqnarray}
\label{eq:Z1}
{\cal Z} = \left(P_{B}P_{C}\right)^{3N/2}{\cal Z}_{\triangle},
\end{eqnarray}
which connects the partition function ${\cal Z}$ of the mixed-spin Ising decorated triangular lattice in the longitudinal magnetic field $h$ with the partition function ${\cal Z}_{\triangle}$ of the simple spin-$1/2$ Ising triangular lattice with the temperature-dependent effective interactions $J_{B}^{\,e\!f\!f}$, $J_{C}^{\,e\!f\!f}$ between the nearest-neighbour\-ing spins and the effective magnetic field $h^{\,e\!f\!f} = h_{B}^{\,e\!f\!f}+h_{C}^{\,e\!f\!f}$ acting on these spins.

It should be emphasized that the relation~(\ref{eq:Z1}) is general, since it is valid for arbitrary values of the decorating spins $s_{B}$, $s_{C}$, model parameters $J_{AB}$, $J_{AC}$, $D_{B}$, $D_{C}$, as well as any values of the external magnetic field $h$. In order to utilize Eq.~(\ref{eq:Z1}) for further rigorous calculations, the exact solution for the partition function ${\cal Z}_{\triangle}$ has to be known. However, the partition function for the spin-$1/2$ Ising triangular lattice has been exactly derived only for the zero magnetic field so far~\cite{Hou50}. This requirement restricts the application of Eq.~(\ref{eq:Z1}) to the case $h^{\,e\!f\!f} = 0$ (or equivalently $\pm h_{B}^{\,e\!f\!f} = \mp h_{C}^{\,e\!f\!f}$), which consequently leads to significant restrictions for model parameters in two possible ways:
\begin{subequations}
\begin{align}
\label{eq:restriction1}
&s_{B} = s_{C} = s\,,\,\, D_{B} = D_{C} = D\,, \,\, \pm J_{AB} = \mp J_{AC} = J\,,
\nonumber\\
&h_{B} = h_{C} = h\\
&\textrm{and}\nonumber\\
\label{eq:restriction2}
&s_{B} = s_{C} = s\,,\,\,  D_{B} = D_{C} = D\,, \,\, J_{AB} = J_{AC} = J\,,
\nonumber\\
&\pm h_{B} = \mp h_{C} = h.
\end{align}
\end{subequations}
Assuming afore-listed conditions, many important physical \linebreak quantities such as the sublattice magnetization $m_A$ and $m_X$ per site of the sublattices $A$ and $X$, respectively, pair correlation functions $c_{A}$, $c_{AX}$ and quadrupolar momenta $q_{X}$ can be exactly calculated by combining the relation~(\ref{eq:Z}) with commonly used exact mapping theorems~\cite{Bar88, Kha90, Bar95} and the differential operator technique~\cite{Hon79, Kan93, Kan97}:
\begin{eqnarray}
\label{eq:mA}
m_{A} \!\!\!&\equiv&\!\!\! \big\langle\sigma_i^z\big\rangle\big|_{h^{\,e\!f\!f} = 0} = \big\langle\sigma_i^z\big\rangle_{\triangle} = m_{\triangle} \,,
\\
\label{eq:cA}
c_{A} \!\!\!&\equiv&\!\!\! \big\langle\sigma_i^z\sigma_{i+1}^z\big\rangle\big|_{h^{\,e\!f\!f} = 0} = \big\langle\sigma_i^z\sigma_{i+1}^z\big\rangle_{\triangle} = c_{\triangle} \,,
\\
\label{eq:mX}
m_{X} \!\!\!&\equiv&\!\!\! \big\langle s_{k}^z\big\rangle\big|_{h^{\,e\!f\!f} = 0}
= K_s(1) + K_s(-1) + 2K_s(0)
{}
\nonumber\\
&& {}
+ 2\alpha m_{\triangle}\big[K_s(1) - K_s(-1)\big] {}
\nonumber\\
&& {}+ 4c_{\triangle}\big[K_s(1)+K_s(-1)-2K_s(0)\big]\,,
\\
\label{eq:cX}
c_{AX} \!\!\!&\equiv&\!\!\! \big\langle \sigma_i^zs_{k}^z\big\rangle\big|_{h^{\,e\!f\!f} = 0} =
\frac{1}{2}\big[K_s(1)-K_s(-1)\big]
{}
\nonumber\\
&& {}
+
2m_{\triangle}\big[K_s(1) + K_s(-1)\big]
+\alpha c_{\triangle}\big[K_s(1)-K_s(-1)\big]\,,
\nonumber\\
\\
\label{eq:qX}
q_{X} \!\!\!&\equiv&\!\!\! \Big\langle \big(s_{k}^z\big)^2\Big\rangle\big|_{h^{\,e\!f\!f} = 0}
= L_s(1) + L_s(-1) + 2L_s(0)
{}
\nonumber\\
&& {}
+ 2\alpha m_{\triangle}\big[L_s(1) - L_s(-1)\big] {}
{}
\nonumber\\
&& {} +4c_{\triangle}\big[L_s(1)+L_s(-1)-2L_s(0)\big]\,.
\end{eqnarray}
We recall that the subscript $X$ in Eqs.~(\ref{eq:mX})--(\ref{eq:qX}) specifies the sublattice formed by decorating spins. Moreover, $X$ also determines the value of the parameter $\alpha$: $\alpha = -2$ if $X=B$, while $\alpha = 2$ if $X=C$. The symbol $\langle\ldots\rangle$ denotes the canonical average performed over the model~(\ref{eq:H_tot}), and $\langle\sigma_i^z\big\rangle_{\triangle} = m_{\triangle}$, $\langle\sigma_i^z\sigma_{j}^z\big\rangle_{\triangle} = c_{\triangle}$ are the magnetization and pair correlation function of the isotropic spin-$1/2$ Ising triangular lattice with the temperature-dependent nearest-neighbour exchange interaction $J^{\,e\!f\!f} = J_{\!B}^{\,e\!f\!f} = J_{\!C}^{\,e\!f\!f}$, respectively. Rigorous analytical solutions of both the quantities $m_{\triangle}$, $c_{\triangle}$ are known~\cite{Pot52, Dom60}, therefore we can restrict calculations to the analytical derivation of the functions $K_s(a)$ and $L_s(a)$:
\begin{eqnarray}
\label{eq:FSGS}
K_s(a) \!\!\!&=&\!\!\! -\dfrac{1}{4}\frac{\displaystyle\sum\limits_{s_{k}^{z}= -s}^s\!\!s_{k}^z\exp{\left[\beta D \big(s_{k}^{z}\big)^{2}\right]}\sinh\left[\beta s_{k}^{z}(a J -h)\right]}{\displaystyle\sum\limits_{s_{k}^{z}= -s}^s\!\!\exp{\left[\beta D \big(s_{k}^{z}\big)^{2}\right]}\cosh\left[\beta s_{k}^{z}(a J -h)\right]}\,,
\\
L_s(a) \!\!\!&=&\!\!\! \dfrac{1}{4}\frac{\displaystyle\sum\limits_{s_{k}^{z}= -s}^s\!\!\big(s_{k}^{z}\big)^{2}\exp{\left[\beta D \big(s_{k}^{z}\big)^{2}\right]}\sinh\left[\beta s_{k}^{z}(a J -h)\right]}{\displaystyle\sum\limits_{s_{k}^{z}= -s}^s\!\!\exp{\left[\beta D \big(s_{k}^{z}\big)^{2}\right]}\cosh\left[\beta s_{k}^{z}(a J -h)\right]}\,.
\end{eqnarray}
In view of above notation, the total magnetization of the studied mixed-spin Ising model normalized per spin can be defined as follows:
\begin{eqnarray}
\label{eq:m+}
m^{+} = \frac{1}{3}\left(m_A + m_B + m_C\right).
\end{eqnarray}
The staggered magnetization per decorating spin can be defined~as:
\begin{eqnarray}
\label{eq:m-}
m^{-} = \frac{1}{2}\left|\,m_B - m_C\right|.
\end{eqnarray}
Other important physical quantities such as Gibbs free energy~${\cal G}$, the entropy ${\cal S}$ and/or the enthalpy ${\cal H}$ can be obtained from fundamental relations of thermodynamics~\cite{Kit58}:
\begin{equation}
\label{eq:GSHCchi}
{\cal G} = -k_{\rm B}T\ln{\cal Z}\,,\quad
{\cal S} = -\left(\frac{\partial{\cal G}}{\partial T}\right)_{h},\quad
{\cal H} = {\cal G} + T{\cal S}.
\end{equation}

Last but not least, critical behaviour of the mixed-spin Ising model defined by the Hamiltonian~(\ref{eq:H_tot}) can also be rigorously examined. It follows from the mapping relation~(\ref{eq:Z1}) that the studied model may exhibit a critical point only if the corresponding spin-$1/2$ Ising triangular lattice is at a critical point. Consequently, the unknown critical temperature $T_c$ of the considered mixed-spin model can be determined from the exact solution for the critical temperature of the isotropic spin-$1/2$ Ising triangular lattice~\cite{Hou50}:
 \begin{equation}
\label{eq:Tc}
\exp{\left(\beta_cJ^{e\!f\!f}\right)} = 3
\quad\Leftrightarrow\quad
W(1)W(-1) - 3W^2(0)\big|_{\beta = \beta_c} = 0.
\end{equation}
Here, $\beta_c = 1/(k_{\rm B}T_c)$ denotes the inverse critical temperature which enters into the functions $W(a) = W_B(\pm a) = W_C(\mp a)$ instead of $\beta$ [see Eq.~(\ref{eq:W})].

\section{Results and discussion}
\label{sec:3}

In this section, we will perform a comprehensive analysis of the most interesting results for the ground state, zero-temperature magnetization process, finite-temperature phase diagrams and MCE of the mixed-spin Ising model defined by the Hamiltonian~(\ref{eq:H_tot}). Before doing so, however, it is worthwhile to remark that the parameter restrictions~(\ref{eq:restriction1}) and~(\ref{eq:restriction2}), which guarantee an exact solvability of the model at finite magnetic fields, give the same results from the physical point of view. Without loss of generality, we can thus limit further discussion to the case~(\ref{eq:restriction1}), which considers an opposite nature (sign) of the exchange interactions between the nodal and decorating spins from two different sublattices $\pm J_{AB} = \mp J_{AC} = J$. Moreover, the condition $J>0$ will also be taken into account due to invariance of the model under the transformation $J\to -J$. To illustrate differences between magnetic properties of the systems whose bonds are decorated by the integer and half-odd-integer spins, we will discuss two particular versions of the mixed-spin Ising model, namely the mixed spin-$(1/2,1)$ model and the mixed spin-$(1/2,3/2)$ model.

\subsection{Ground-state phase diagrams}
\label{subsec:31}

\begin{figure*}[t!]
\vspace{0.1cm}
\centering
  \includegraphics[width=0.46\textwidth]{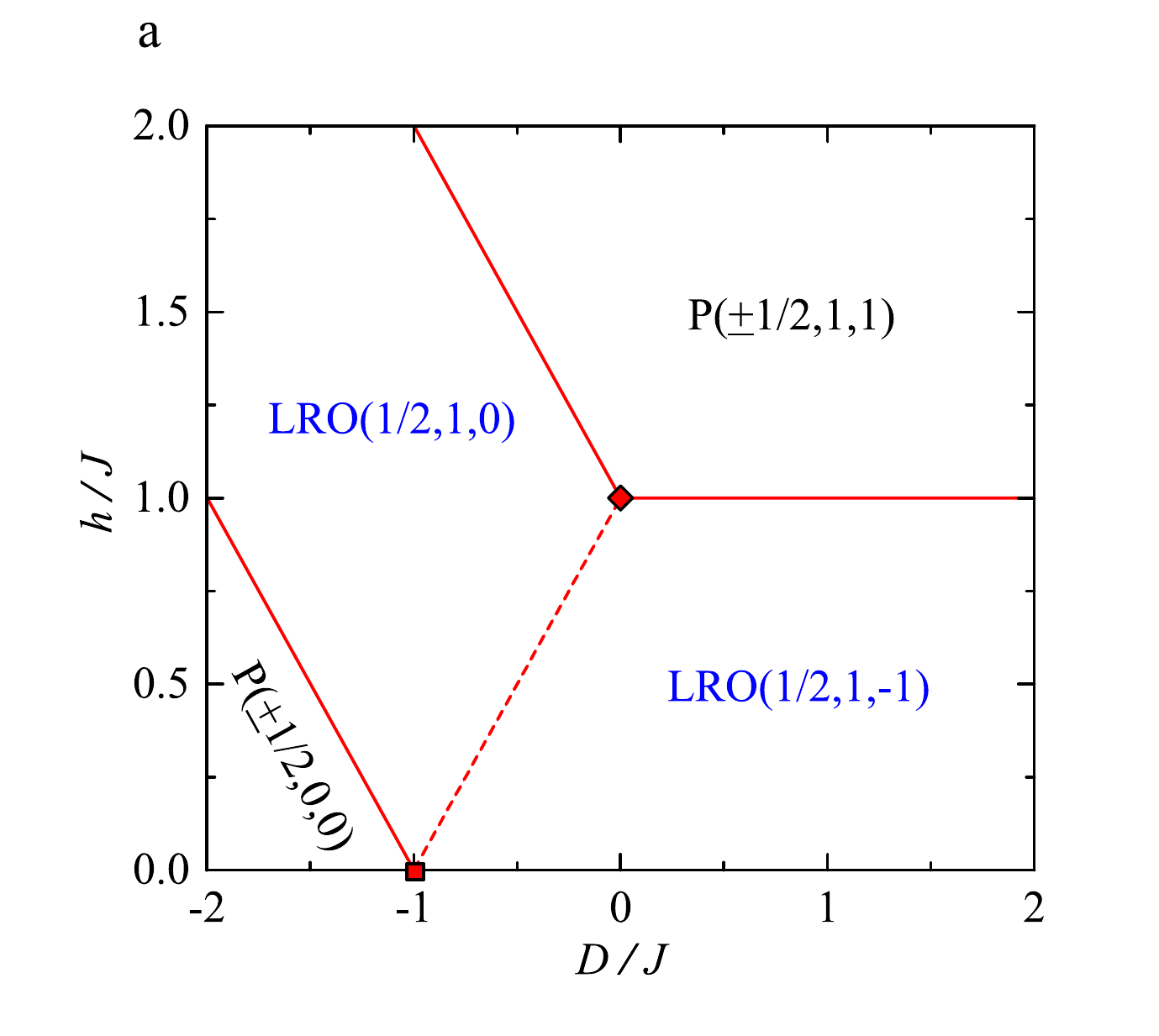}
  \includegraphics[width=0.46\textwidth]{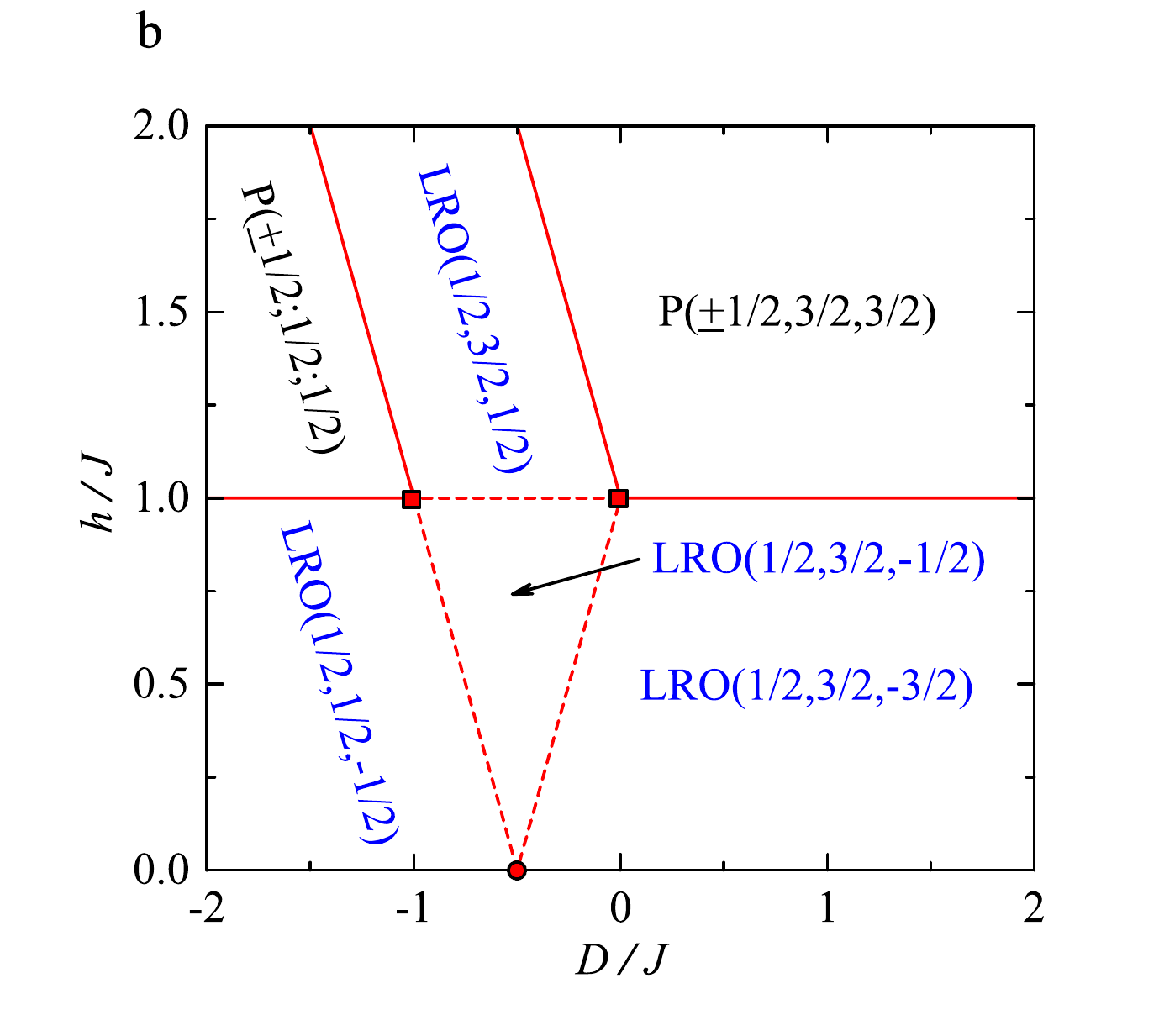}\\
  \includegraphics[width=0.96\textwidth]{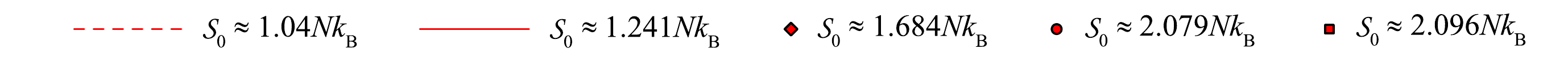}
\vspace{-0.25cm}
\caption{(Color online) Ground-state phase diagrams in the $D-h$ plane for the mixed spin-$(1/2,1)$ model (figure~a) and for the mixed spin-$(1/2,3/2)$ model (figure~b). The individual phases are distinguished by the spin states of distinct sublattices listed in rounded brackets in order $(A, B, C)$ or $(A, C, B)$ depending on whether the condition $J_{AB} = -J_{AC} = J>0$ or $-J_{AB} = J_{AC} = J>0$ is valid, respectively. Legend states the values of zero-temperature residual entropy at individual phase boundaries and higher-order coexistence points.}
\label{fig:2}
\end{figure*}

We start by analyzing possible spin arrangements which appear in the ground state. For this purpose, two ground-state diagrams constructed the $D-h$ plane are depicted in Fig.~\ref{fig:2} for both afore-listed spin versions of the model. Solid and dashed lines in displayed phase diagrams indicate discontinuous (first-order) phase transitions between neighbouring ground states. Full diamond, circle and squares denote crossing points of the transition lines, at which three or four phases coexist together with other possible spin microstates (for further details see the discussion below).

As one can see from Fig.~\ref{fig:2}, depicted lines divide the $D-h$ plane into several regions which correspond to various long-range ordered (LRO) and paramagnetic (P) ground states. The individual phases are distinguished according to spin states of the sublattices $A$, $B$ and $C$, which are listed in rounded brackets. Clearly, the LRO (as well as P) phases differ from each other only by respective spin arrangements of the decorating spins from $B$- and $C$-sublattices. In each LRO phase, nodal spins of the $A$-sublattice  reside the spin state $\sigma_i^{z} = 1/2$, while in each P phase, these spins may occupy one of two possible spin states $\sigma_i^{z} = 1/2$ or $-1/2$ with the same probability. It is worthy to note that spin frustration of the sublattice $A$ observed in the latter group of phases is a result of the competition between ferromagnetic and antiferromagnetic exchange interactions connecting nodal spins with the nearest-neighbouring decorating spins from $B$- and $C$-sublattices. If the decorating spins $s=1$ constitute $B$- and $C$-sublattices, the spin frustration of the sublattice $A$ is present in the ground state for weak magnetic fields $0\leq h/J < -1 - D/J$ (if $D/J<-1$), as well as, for strong magnetic fields $h/J > 1 - D/J$ (if $D/J<0$) and $h/J > 1$ (if $D\geq0$), as illustrated in Fig.~\ref{fig:2}a. By contrast, if the spins $s=3/2$ occupy $B$- and $C$-sites of the decorated triangular lattice, then, the sublattice $A$ becomes frustrated only for magnetic fields which are stronger than the nearest-neighbour interaction $J$, namely, $1<h/J<-1-2D/J$, $h/J>1-2D/J$ (if $D/J<0$) and $h/J > 1$ (if $D\geq0$), as shown in Fig.~\ref{fig:2}b. Macroscopic degeneracies of the P phases are proportional to a total number of the $A$-sites regardless of the magnitude of decorating spins. This fact is reflected in the same residual entropy ${\cal S}_{0} = Nk_{\rm B}\ln 2\approx 0.693Nk_{\rm B}$ observed for these phases.

A high macroscopic degeneracy can also be found at the phase boundaries separating individual ground states and at \linebreak crossing points of different ground-state phase boundaries. The degeneracy at these particular regions are apparent from zero-temperature values of the residual entropy listed in the legend of Fig.~\ref{fig:2}. Obviously, the residual entropy reaches the value \linebreak ${\cal S}_{0} \approx 1.04Nk_{\rm B}$ at the boundaries separating different LRO phas\-es. This finite value points to the macroscopic degeneracy of the size $2^{3N/2}$, which can be attributed to zero-temperature fluctuations of the spins either from $B$- or $C$-sublattice between two spin states found in neighbouring LRO phases. The residual entropy at other phase boundaries is slightly higher, namely, ${\cal S}_{0} \approx 1.241Nk_{\rm B}$. This non-trivial value includes two contributions ${\cal S}_{01} \approx 1.04Nk_{\rm B}$ and ${\cal S}_{02} \approx 0.201Nk_{\rm B}$. The larger entropy contribution comes from zero-temperature spin fluctuations in the sublattice $B$ or $C$, while the smaller one is associated with spin fluctuations in the $A$-sublattice. The sublattice $A$ is fully saturated in the LRO phase, but fully frustrated in the P phase. The crossing points of different ground-state phase boundaries also exhibit macroscopic degeneracies, which manifest themselves in three different residual entropies ${\cal S}_{0} \approx 1.684Nk_{\rm B}$, $2.079Nk_{\rm B}$ and $2.096Nk_{\rm B}$. The lowest value ${\cal S}_{0} \approx 1.684Nk_{\rm B}$ corresponds to the crossing point with the coordinates~$\left[D/J, h/J\right]=\left[0, 1\right]$, which appears in the zero-temperature diagram of the mixed spin-$(1/2,1)$ Ising model (see Fig.~\ref{fig:2}a). This point is a triple point at which spin configurations of the neighbouring phases ${\rm P}(\pm1/2, 1,1)$, ${\rm LRO}(1/2, 1,0)$, ${\rm LRO}(-1/2, 1,-1)$ coexist with another two three-spin configurations $(-1/2, -1,1)$, $(-1/2, 0,1)$. One can easily prove that the enthalpy of both the aforementioned ground states normalized per one nodal lattice site acquire the same value as the enthalpy of the observed microstates if $D/J=0$ and $h/J=1$. The second value of the residual entropy ${\cal S}_{0} \approx 2.079Nk_{\rm B}$ can be detected at the triple point~$\left[D/J, h/J\right]=\left[-0.5, 0\right]$, which occurs in the ground-state phase diagram of the mixed spin-$(1/2,3/2)$ Ising model (see Fig.~\ref{fig:2}b). Spin configurations of three neighbouring phases ${\rm LRO}(1/2, 1/2,-1/2)$, ${\rm LRO}(1/2, 3/2,-1/2)$, ${\rm LRO}(1/2, 3/2,-3/2)$ and three spin microstates $(-1/2, -1/2,1/2)$, $(-1/2, -3/2,1/2)$, $(-1/2, -3/2, 3/2)$ are in a thermodynamic equilibrium at this particular point. 
Finally, the highest value of the residual entropy ${\cal S}_{0} \approx 2.096Nk_{\rm B}$ can be reached in the ground states of both the studied Ising models, namely at the triple point \linebreak $\left[D/J, h/J\right]=\left[-1, 0\right]$ appearing in the ground-state phase diagram of the mixed spin-$(1/2,1)$ model (see Fig.~\ref{fig:2}a), and at two quadruple points $\left[D/J, h/J\right]=\left[-1, 1\right]$, $\left[D/J, h/J\right]=\left[0, 1\right]$ found in the ground-state phase diagram of the mixed spin-$(1/2,3/2)$ model (see Fig.~\ref{fig:2}b). This unusually high entropy is a result of the coexistence of up to seven distinct types  of spin configurations. To be more specific, the triple point $\left[-1, 0\right]$ in Fig.~\ref{fig:2}a is characterized by the coexistence of the phases ${\rm P}(\pm1/2, 0,0)$, ${\rm LRO}(1/2, 1,0)$, ${\rm LRO}(1/2, 1,-1)$  with the three-spin microstates $(1/2, 0,-1)$, $(-1/2, -1,0)$, $(-1/2, -1, 1)$, \linebreak $(-1/2, 0, 1)$. At the quadruple point $\left[-1, 1\right]$ in Fig.~\ref{fig:2}b, the phases ${\rm P}(\pm1/2, 1/2,1/2)$, ${\rm LRO}(1/2, 1/2,-1/2)$, ${\rm LRO}(1/2, 3/2,1/2)$, \linebreak ${\rm LRO}(1/2, 3/2,-1/2)$ are in thermodyna\-mic equilibrium with the microstates $(-1/2, -1/2,1/2)$, $(-1/2, 1/2, 3/2)$, \linebreak $(-1/2, -1/2, 3/2)$. Finally, the quadruple point $\left[0, 1\right]$ in the same figure is characterized by the thermodynamic equilibrium of the phases ${\rm P}(\pm1/2, 3/2,3/2)$, ${\rm LRO}(1/2, 3/2,1/2)$, \linebreak ${\rm LRO}(1/2, 3/2,-1/2)$,  ${\rm LRO}(1/2, 3/2,-3/2)$ and the microstates $(-1/2, -3/2,3/2)$, $(-1/2, -1/2, 3/2)$, $(-1/2, -1/2, 3/2)$.

\subsection{Zero-temperature magnetization process}
\label{subsec:32}

\begin{figure*}[t!]
\vspace{0.1cm}
\centering
  \includegraphics[width=0.46\textwidth]{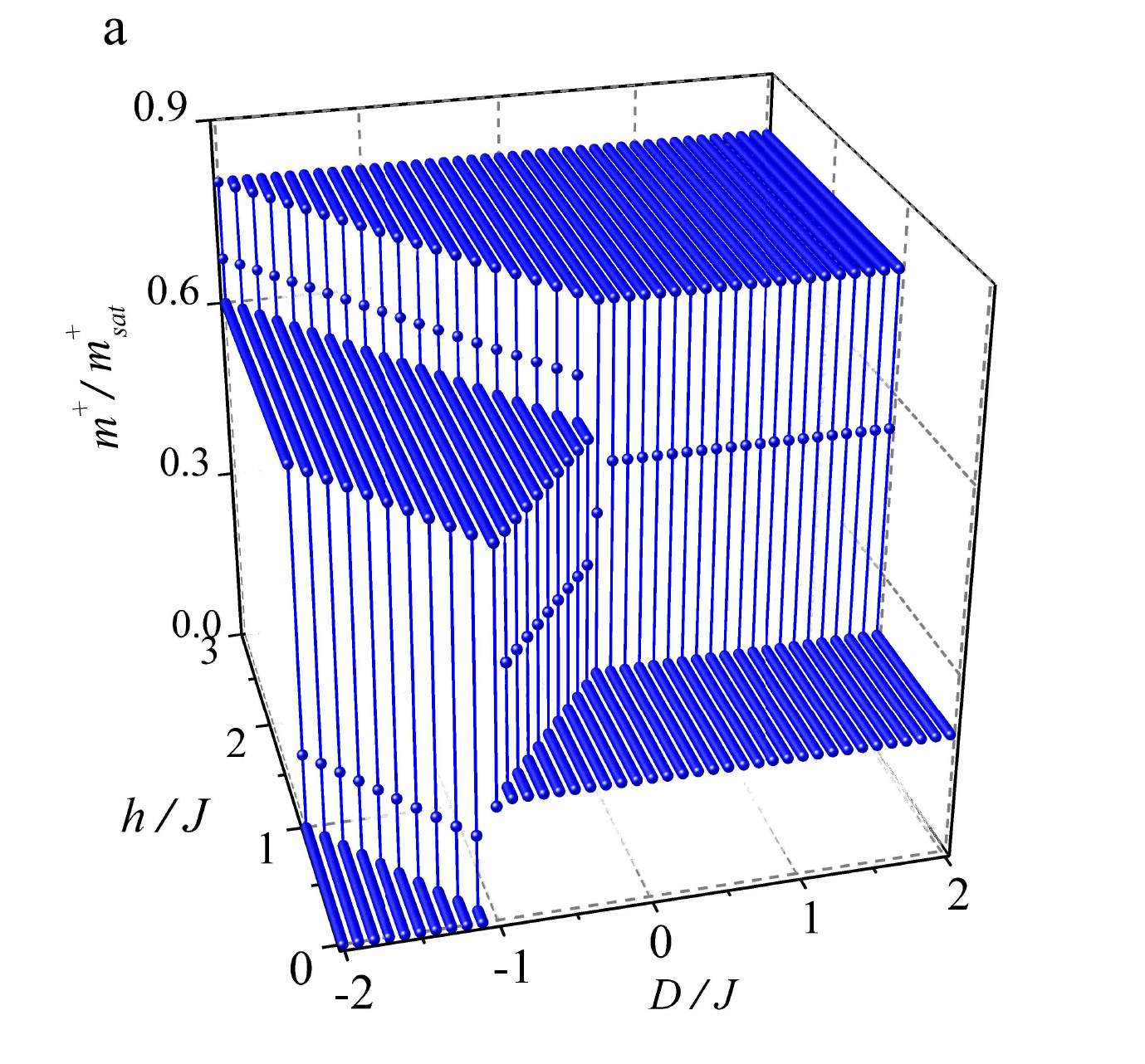}
  \hspace{-0.25cm}
  \includegraphics[width=0.46\textwidth]{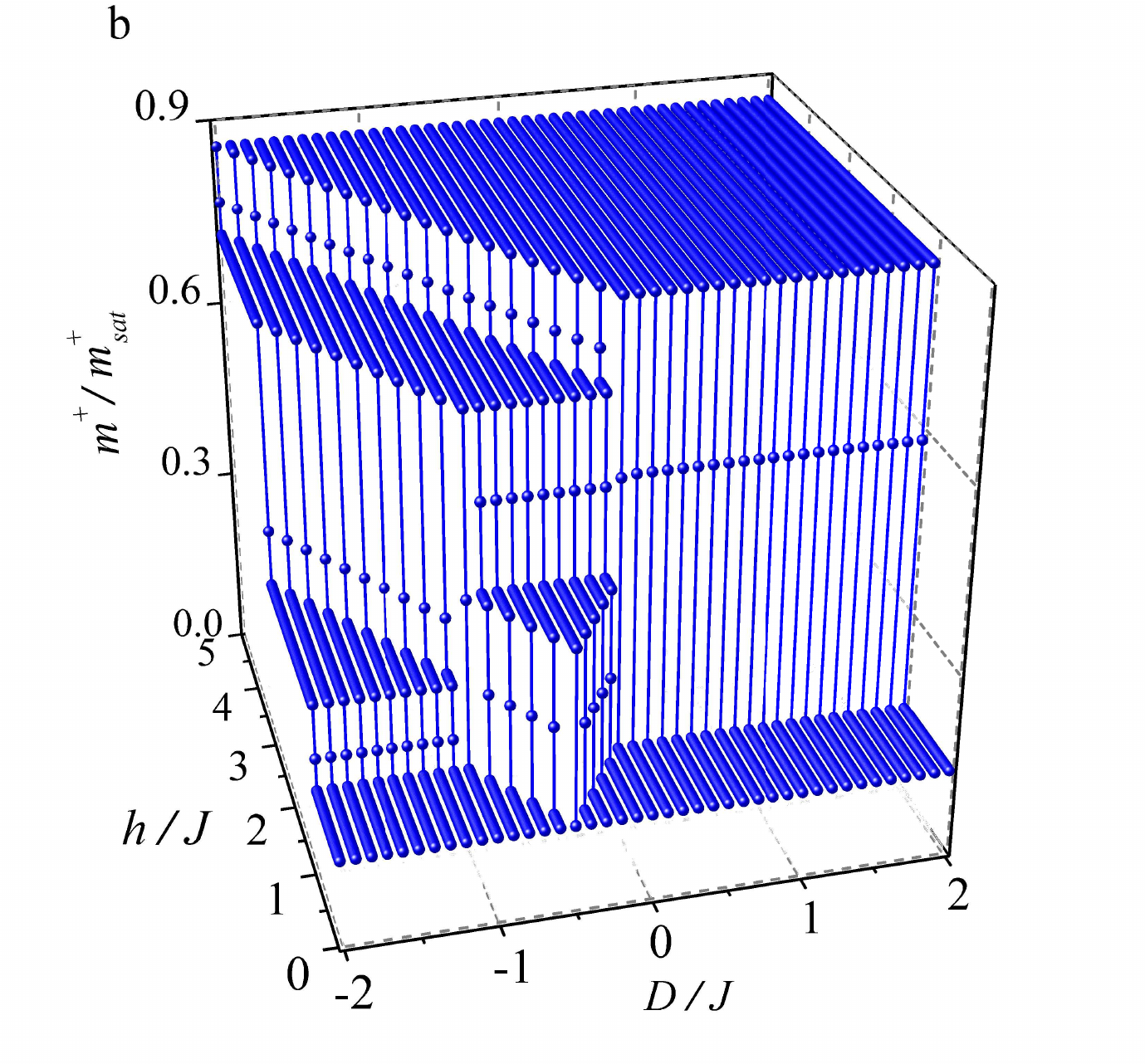}
\vspace{-0.25cm}
\caption{(Color online) 3D plot of the zero-temperature magnetization reduced to its saturation value as a function of the magnetic field and single-ion anisotropy for the mixed spin-$(1/2,1)$ Ising model (figure a) and the mixed spin-$(1/2,3/2)$ Ising model (figure b).
}
\label{fig:3}
\end{figure*}

Ground-state phase diagrams in Fig.~\ref{fig:2} suggest a rich diversity of magnetization processes, which depend on the magnitude of decorating spins and the mutual interplay between the model parameters $D$ and $J$. Figure~\ref{fig:3} presents three-dimensional (3D) images of the zero-temperature magnetization processes for the studied versions of the 2D mixed-spin Ising model in order to prove this claim.

It is clear from Fig.~\ref{fig:3}a that the magnetization curves corresponding to the mixed spin-$(1/2,1)$ Ising model include in total four different plateaus, namely at the zero magnetization, one-fifth, three-fifths and four-fifths of the saturation magnetization. Referring to the ground-state phase diagram plotted in Fig.~\ref{fig:2}a one can conclude that the zero magnetization corresponds to the disordered phase ${\rm P}(\pm1/2, 0,0)$ and other magnetization plateaus observed at $m^+/m_{sat}= 1/5$, $3/5$, $4/5$ reflect the existence of the phases ${\rm LRO}(1/2, 1,-1)$, ${\rm LRO}(1/2, 1,0)$, ${\rm P}(\pm1/2, 1,1)$, respectively.

More complex magnetization scenario can be found for the mixed spin-$(1/2,3/2)$ Ising model. Namely, if decorating lattice sites are occupied by the spins of the magnitude $s=3/2$, the zero-temperature magnetization curves include plateaus at one-, two-, three-, five- and six-sevenths of the saturation magnetization (see Fig.~\ref{fig:3}b). It can be immediately deduced from a comparison of Fig.~\ref{fig:3}b with Fig.~\ref{fig:2}b that the lowest magnetization plateau appearing at $m^+/m_{sat}=1/7$ corresponds to either the phase ${\rm LRO}(1/2, 1/2, -1/2)$ or ${\rm LRO}(1/2, 3/2, -3/2)$. Another three magnetization plateaus at $m^+/m_{sat}=2/7$, $3/7$ and $5/7$ are pertinent to the phases ${\rm P}(\pm1/2, 1/2, 1/2)$, \linebreak ${\rm LRO}(1/2, 3/2,-1/2)$ and ${\rm LRO}(1/2, 3/2, 1/2)$, respectively. Finally, the highest plateau identified at $m^+/m_{sat}= 6/7$ is a result of the phase ${\rm P}(\pm1/2, 3/2, 3/2)$.

Interestingly, the displayed zero-temperature magnetization curves of both the studied mixed-spin models never reach the saturation magnetization ($m_{sat} = 5/6$ for the spin case $s=1$ and $m_{sat} = 7/6$ for the spin case $s=3/2$). This behaviour can be attributed to the fact that all three sublattices $A$, $B$ and $C$ are never simultaneously fully polarized into the magnetic-field direction in the zero-temperature $D-h$ plane due to a competition between exchange interactions connecting nodal spins of the $A$-sublattice with the nearest-neighbouring spins from the $B$- and $C$-sublattices. Moreover, discontinuous magnetization jumps correspond to the discontinuous field-induced phase transitions between individual ground states, which manifest themselves as intermediate magnetization plateaus. At relevant critical fields, the total magnetization $m^+$ takes non-trivial values, which are determined by the magnitude of decorating spins and spin arrangements in neighbouring phases.

\subsection{Finite-temperature behaviour}
\label{subsec:33}

\begin{figure*}[t!]
\vspace{-1.0cm}
\centering
  \includegraphics[width=0.46\textwidth]{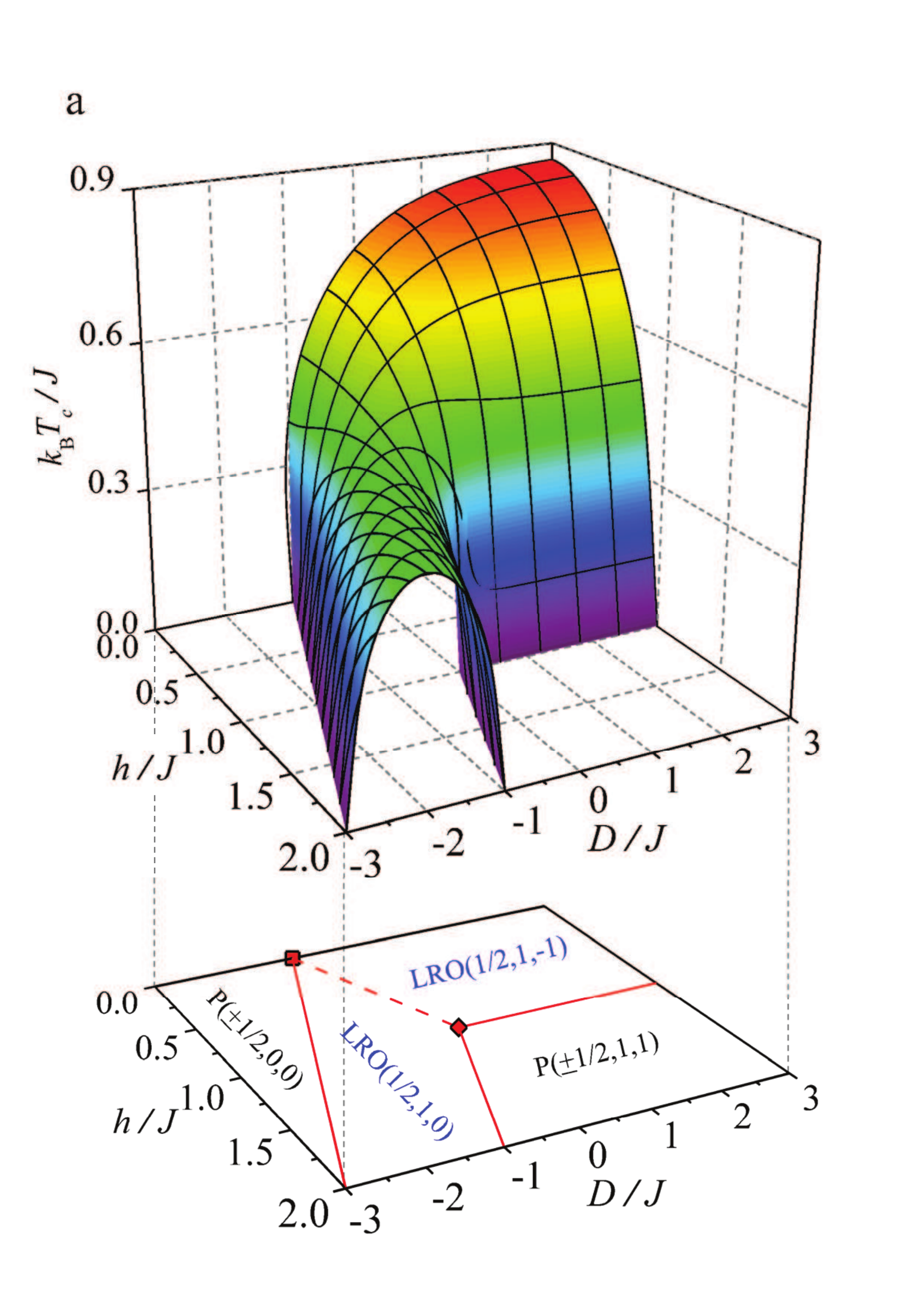}
  \hspace{-0.25cm}
  \includegraphics[width=0.46\textwidth]{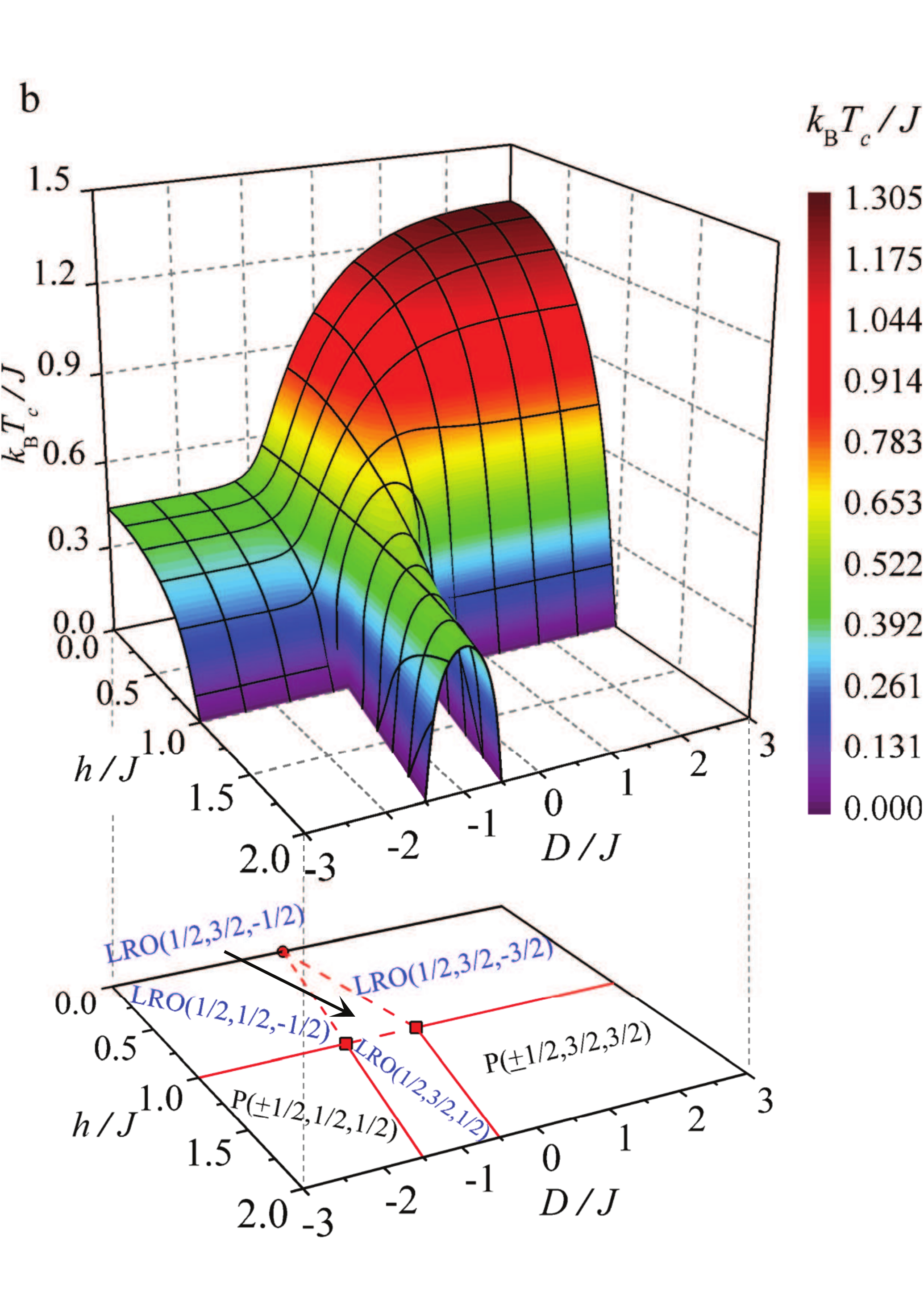}
\vspace{-0.5cm}
\caption{(Color online) Global finite-temperature phase diagrams supplemented with shifted zero-temperature $D-h$ planes showing appropriate ground-state phase diagrams for the mixed spin-$(1/2,1)$ Ising model  (figure a) and the mixed spin-$(1/2,3/2)$ Ising model (figure b).
}
\label{fig:4}
\vspace{8mm}
\centering
  \includegraphics[width=0.46\textwidth]{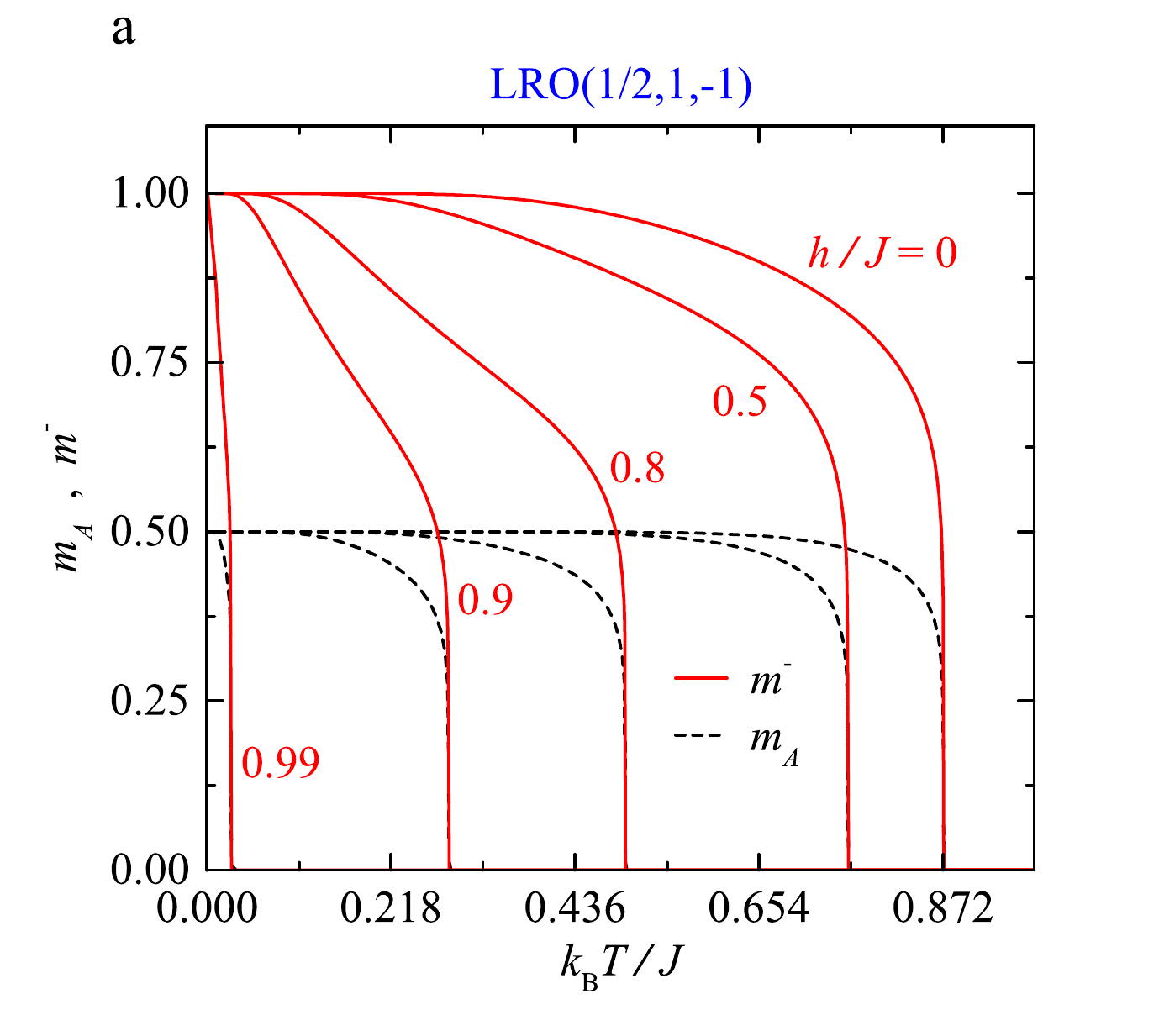}
    \hspace{-0.25cm}
  \includegraphics[width=0.46\textwidth]{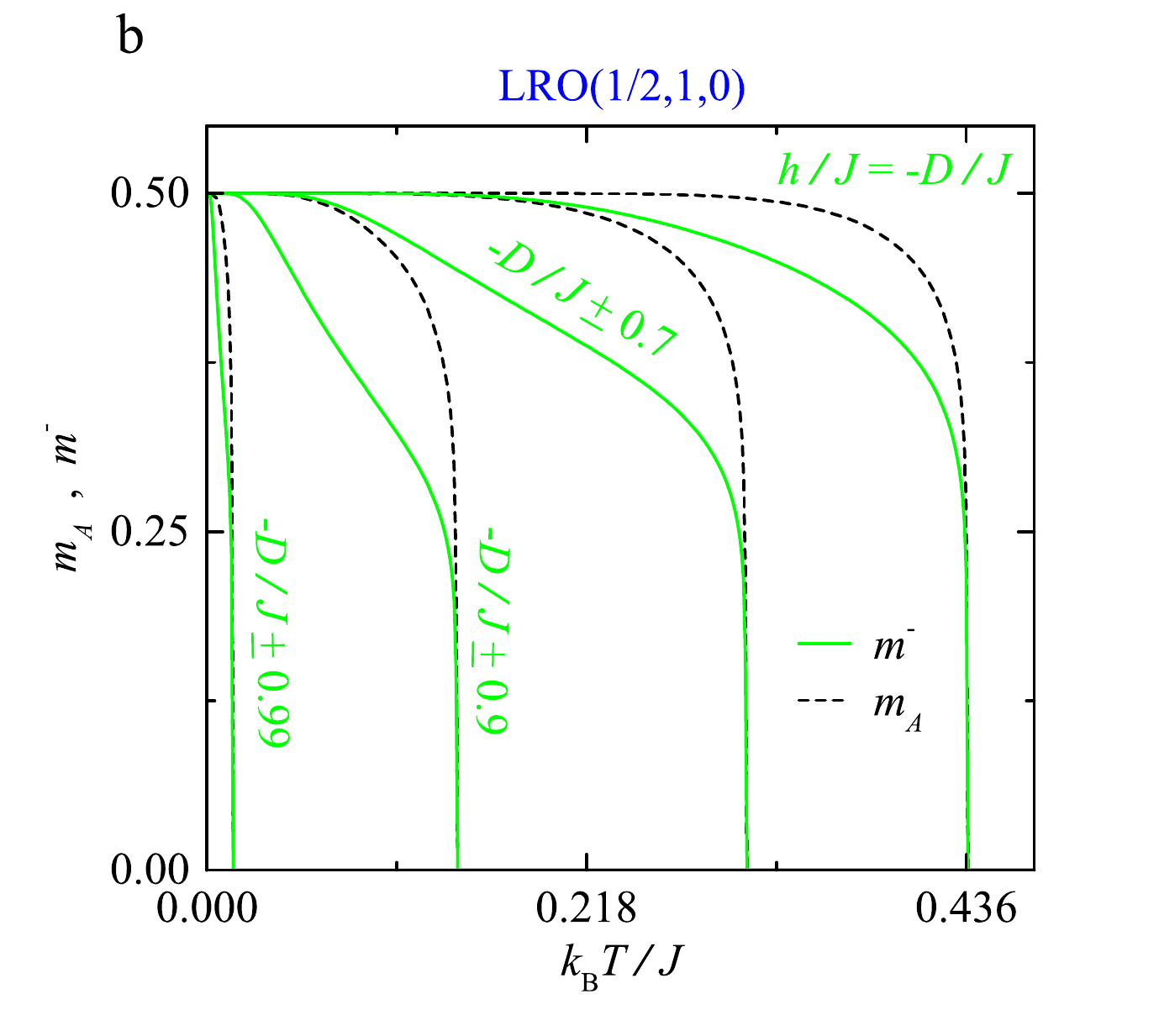}
\vspace{-0.25cm}
\caption{(Color online) Temperature dependencies of the sublattice magnetization $m_A$ and the staggered magnetization $m^{-}$ for the mixed spin-$(1/2,1)$ Ising model. The~displayed curves correspond to the phases LRO$(1/2,1,-1)$ for $D/J\to\infty$ (figure~a) and LRO$(1/2,1,0)$ for $D/J\to-\infty$ (figure~b).
}
\label{fig:5}
\end{figure*}

In order to determine stability regions of individual LRO phases at finite temperatures, two global finite-temperature \linebreak phase diagrams are illustrated in Fig.~\ref{fig:4}. The figure provides 3D views of the critical behaviour of the mixed spin-$(1/2,1)$ and spin-$(1/2,3/2)$ Ising models in dependence on the single-ion anisotropy parameter $D$ and the external magnetic field $h$. All points of the 3D surfaces are unique solutions of the condition~(\ref{eq:Tc}), and therefore they represent points of continuous (second-order) phase transitions separating the LRO phase from the disordered P one. For better clarity of the finite-temperature behaviour, the displayed global phase diagrams are completed with shifted zero-temperature $D-h$ planes showing appropriate ground-state phase diagrams of the investigated mixed-spin Ising model.

Figure~\ref{fig:4}a shows the critical temperature of the mixed spin-$(1/2,1)$ Ising model. Obviously, the phase ${\rm LRO}(1/2, 1,-1)$, appearing at zero temperature for relatively weak magnetic fields $0 \leq h/J < 1 + D/J$ (if $-1<D/J<0$) and $0\leq h/J < 1$ (if $D/J>0$), is thermally more stable than the phase ${\rm LRO}(1/2, 1,0)$, which can be observed in the field ranges $-1-D/J < h/J < 1 - D/J$ (if $D/J<-1$) and $1+D/J < h/J < 1 - D/J$ (if $-1<D/J<0$). In fact, the highest critical temperature $k_{\rm B}T_c^{max}/J = 2\left[\ln\left(5+2\sqrt{6}\right)\right]^{-1}\!\!\approx 0.872$ of the phase ${\rm LRO}(1/2, 1,-1)$ achie\-ved at the zero magnetic field in the limit $D/J\to\infty$ is twice as large as the highest critical temperature of the phase \linebreak ${\rm LRO}(1/2, 1,0)$, which can be detected along the line $h/J=-D/J$ for sufficiently strong but finite fields $h/J\gg 1$ and in the asymptotic limit $h/J\to\infty$. The maximum possible critical temperature of the latter phase is $k_{\rm B}T_c^{max}/J = \left[\ln\left(5+2\sqrt{6}\right)\right]^{-1}\!\!\approx 0.436$, because only a half of all decorating spins in the system are in magnetic states unlike ${\rm LRO}(1/2, 1,-1)$. It is quite obvious from Fig.~\ref{fig:4}a that critical temperatures of both LRO phases quickly drop to the zero at ground-state phase boundaries separating LRO and P phases.

Qualitatively similar behaviour can be observed for the \linebreak mixed spin-$(1/2,3/2)$ Ising model. It can be concluded from Fig.~\ref{fig:4}b that the most thermally stable phase is \linebreak ${\rm LRO}(1/2, 3/2,-3/2)$, which appears in the ground state at relatively low magnetic fields $0 \leq h/J < 1 + 2D/J$ (if $-0.5<D/J<0$) and $0\leq h/J < 1$ (if $D/J>0$). The highest critical temperature of this phase $k_{\rm B}T_c^{max}/J = 3\left[\ln\left(5+2\sqrt{6}\right)\right]^{-1}\!\!\approx 1.308$ can be reached in the asymptotic limit $D/J\to\infty$ at $h/J=0$. On the other hand, critical temperatures corresponding to other two phases ${\rm LRO}(1/2,1/2,1/2)$ and ${\rm LRO}(1/2,3/2,1/2)$ achieve their maximum values at one-third of the aforementioned one on account of a reduction of the zero-temperature staggered magnetization of decorating spins to one-third of that corresponding to ${\rm LRO}(1/2, 3/2,-3/2)$. For the former phase ${\rm LRO}(1/2,1/2,1/2)$, which is stable at the magnetic fields $0 < h/J < 1$ (if $D/J<-1$) and $0 < h/J < -1 - 2D/J$ (if $-1<D/J<-0.5$), the highest critical temperature $k_{\rm B}T_c^{max}/J \approx 0.436$ can be observed at $h=0$ in the limit $D/J\to-\infty$. For the latter phase ${\rm LRO}(1/2,3/2,1/2)$, emerging at the magnetic fields $-1-2D/J < h/J < 1-2D/J$ (if $D/J<-1$) and $1 < h/J < 1 - 2D/J$ (if $-1<D/J<0$), the critical temperature reaches its maximum value $k_{\rm B}T_c^{max}/J \approx 0.436$ along the line $h/J=-2D/J$ if $D/J\ll-1$ and/or $D/J\to-\infty$. As expected, critical temperatures of all three LRO phases sharply drop down to the zero when approaching the ground-state phase boundaries LRO--P (see Fig.~\ref{fig:4}b).

\begin{figure}[t!]
\vspace{0.25cm}
\centering
  \includegraphics[width=0.96\columnwidth]{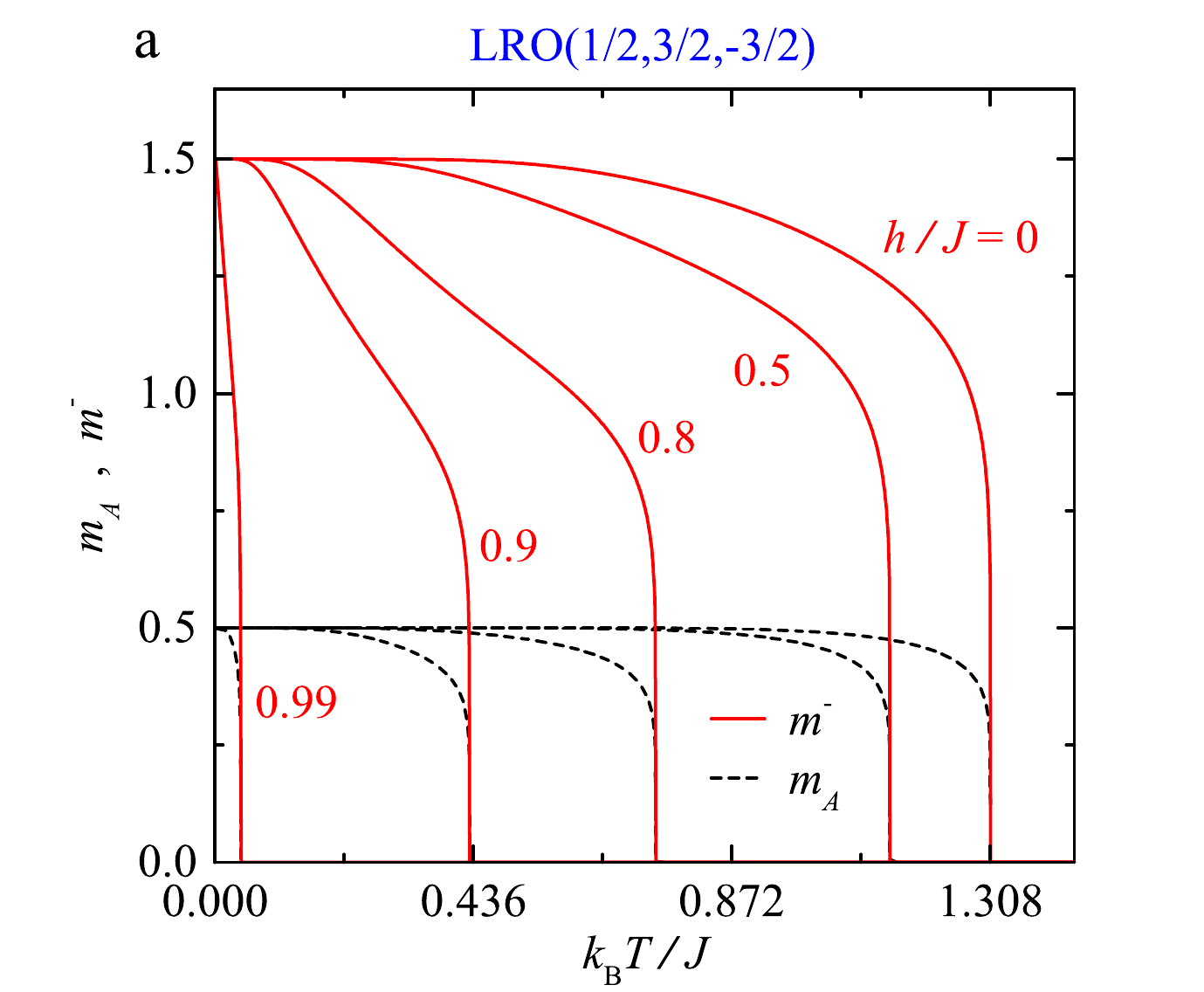}\vspace{1mm}
  \includegraphics[width=0.96\columnwidth]{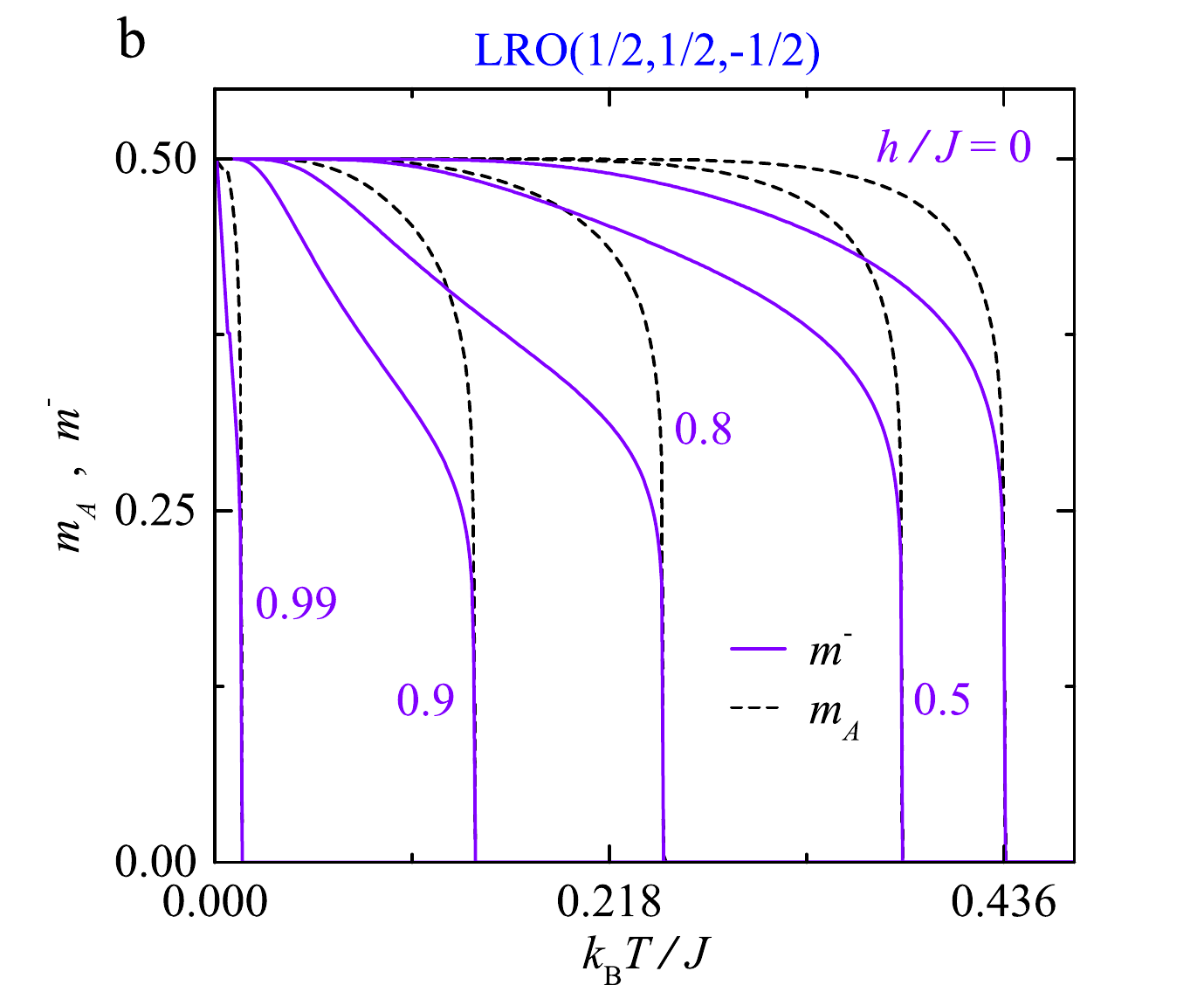}\vspace{1mm}
  \includegraphics[width=0.96\columnwidth]{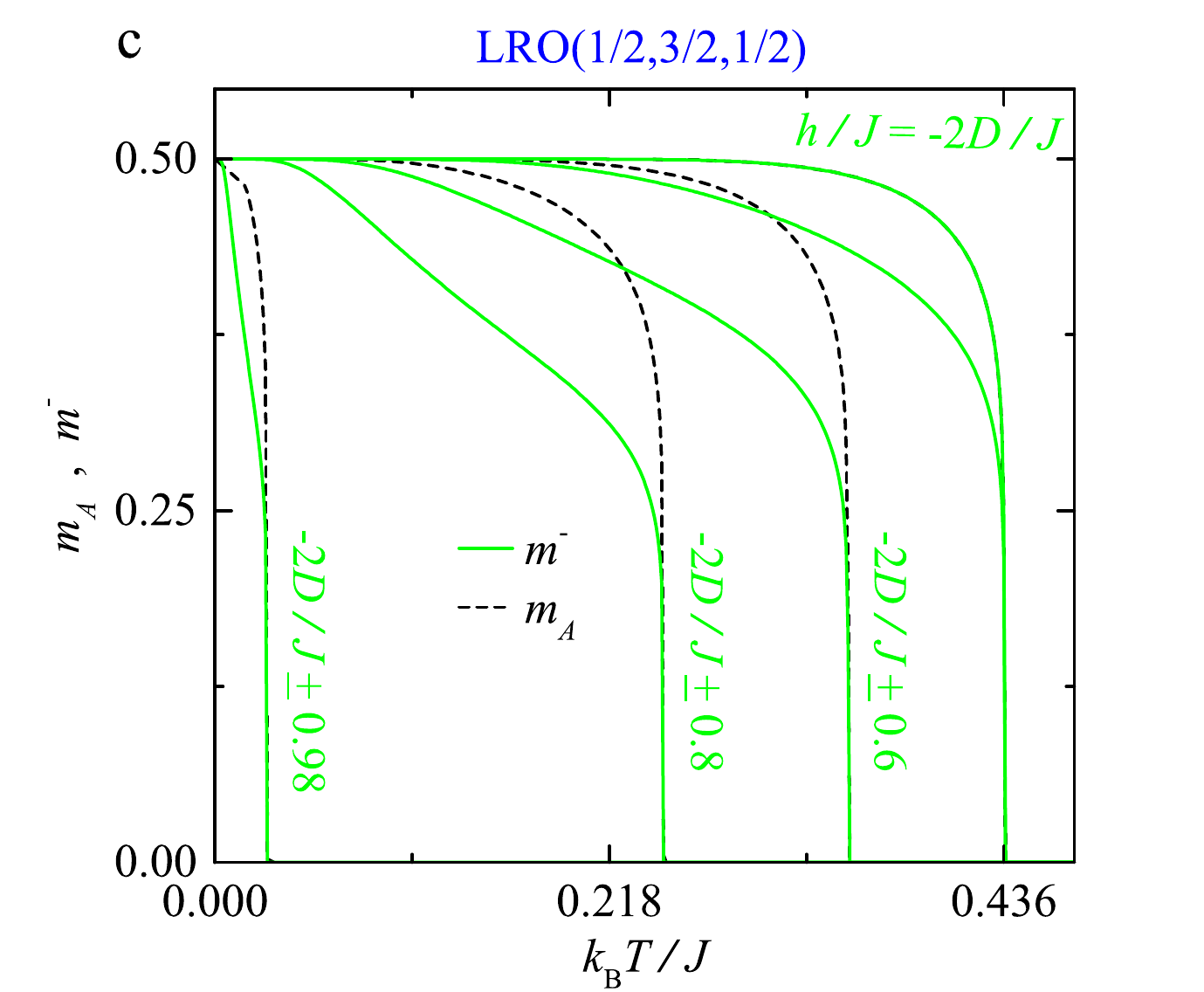}
\vspace{-0.15cm}
\caption{(Color online) Temperature dependencies of the sublattice magnetization $m_A$ and the staggered magnetization $m^{-}$ for the mixed spin-$(1/2,3/2)$ Ising model. The displayed curves correspond to the phases LRO$(1/2,3/2,-3/2)$ for $D/J\to\infty$ (figure a), LRO$(1/2,1/2,-1/2)$ for $D/J\to-\infty$ (figure b) and LRO$(1/2,3/2,1/2)$ for $D/J\to-\infty$.
}
\label{fig:6}
\vspace{-0.75cm}
\end{figure}

The above findings can be independently confirmed by temperature dependencies of the sublattice magnetization $m_A$ corresponding to nodal spins (dashed lines) and the staggered magnetization $m^{-}$ of the decorating spins given by Eq.~(\ref{eq:m-}) (solid lines), which are plotted in Figs.~\ref{fig:5} and~\ref{fig:6}. In both the figures, the values of the parameters $D$, $h$, $J$ are chosen so that the displayed magnetization curves achieve the highest possible critical temperatures of the LRO phases observed for corresponding versions of the mixed-spin Ising model. Obviously, the magnetization $m_A$ and $m^{-}$ exhibit qualitatively the same thermal trends regardless of the magnitude of decorating spins. These trends can be described by the extended N\'eel's classification~\cite{Nee48,Chi97,Str06}. The sublattice magnetization $m_A$ exhibits solely familiar Q-type dependencies, which are characterized by a steep decrease just in the vicinity of the critical temperature. On the other hand, the staggered magnetization $m^{-}$ decreases to zero with the increasing temperature the faster, the closer the parameters $D$ and $h$ are selected to the ground-state boundaries LRO--P. As a result, the temperature dependencies of $m^{-}$ may change from the standard Q-type dependence to a more interesting R-type dependence, which exhibits a relatively rapid decline at moderate temperatures before a sharp drop to the zero value is reached at the critical point.

In accordance with the ground-state analysis, the staggered magnetization of the mixed spin-$(1/2,1)$ Ising model starts from the values $m^{-}_{0} = 1$, if the phase ${\rm LRO}(1/2, 1, -1)$ constitutes the ground state, and $m^{-}_{0} = 1/2$, if ${\rm LRO}(1/2, 1, 0)$ is stable at zero temperature. The temperature dependencies of $m^{-}$ and $m_A$ plotted in Fig.~\ref{fig:5}a for the infinitely strong easy-axis anisotropy and the zero magnetic field sharply fall down to the zero at the temperature $k_{\rm B}T_c/J \approx 0.872$, which relates to the previously discussed critical behaviour of the phase ${\rm LRO}(1/2, 1, -1)$ at $h/J=0$ and for $D/J\to\infty$. On the other hand, the $m^{-}(T)$ and $m_A(T)$ curves corresponding to the phase ${\rm LRO}(1/2, 1, 0)$ drop to the zero value at a half critical temperature $k_{\rm B}T_c/J \approx 0.436$ if a combination of the parameters $h/J = -D/J$, $D/J \to-\infty$ is assumed (see Fig.~\ref{fig:5}b).

Similar conclusions can be reached from temperature dependencies of the sublattice magnetization $m_A$ and the staggered magnetization $m^{-}$ of the mixed spin-$(1/2, 3/2)$ Ising mo\-del, which are illustrated in Fig.~\ref{fig:6}. Namely, temperature variations of the staggered magnetization start either from the value $m^{-}_{0} = 3/2$ or from three times lower value $m^{-}_{0} =1/2$ depending on whether the phase ${\rm LRO}(1/2, 3/2, -3/2)$ or the phases ${\rm LRO}(1/2, 1/2, -1/2)$, ${\rm LRO}(1/2, 3/2, 1/2)$ constitute the ground state. The $m^{-}(T)$ and $m_A(T)$ curves corresponding to the phase ${\rm LRO}(1/2, 3/2, -3/2)$ sharply drop to zero at the maximum critical temperature $k_{\rm B}T_c/J \approx 1.308$ if $h/J = 0$ and $D/J\to\infty$ (see Fig.~\ref{fig:6}a), while for two other phases ${\rm LRO}(1/2, 1/2, -1/2)$ and ${\rm LRO}(1/2, 3/2, 1/2)$, they fall down to the zero value at three times lower critical temperature $k_{\rm B}T_c/J \approx 0.436$ if $h/J = 0$ and $h/J=-2D/J$, respectively (see Figs.~\ref{fig:6}b, c). We note that the asymptotic limit of the single-ion anisotropy parameter $D/J\to-\infty$ has to be assumed in both aforementioned LRO phases. The observed maximum critical temperatures are in agreement with the previously discussed asymptotic critical behaviour of these phases. The closer the values of $D$ and $h$ are taken to the LRO--P phase boundaries, the faster is the decline of the staggered magnetization $m^{-}$ before a sharp drop to the zero can be detected. On the other hand, the sublattice magnetization $m_A$ always exhibits a steep decrease only in a vicinity of the critical temperature regardless of the relevant spin arrangement of individual LRO phases.

\subsection{Isothermal magnetic entropy change}
\label{subsec:34}

To characterize magnetocaloric properties of the model during the isothermal process, the isothermal magnetic entropy change~$\Delta {\cal S}_{iso}$ will be examined. This quantity can be calculated as a difference of the magnetic entropies at a finite and zero magnetic field under the fixed temperature $T$~\cite{Tis03}:
\begin{equation}
\label{eq:S_iso}
\Delta{\cal S}_{iso} (T, \Delta h) = {\cal S}(T, h>0) - {\cal S}(T, h=0).
\end{equation}
In the presented convention, $-\Delta {\cal S}_{iso}>0$ corresponds to a conventional MCE, while $-\Delta {\cal S}_{iso}<0$ points to an inverse MCE. In general, the enhanced conventional (inverse) MCE may be expected in a vicinity of field- or temperature-induced phase transitions, because two or more phases are in a thermodynamic equilibrium at relevant critical points.
\begin{figure*}[t!]
\vspace{0.1cm}
\centering
  \includegraphics[width=0.46\textwidth]{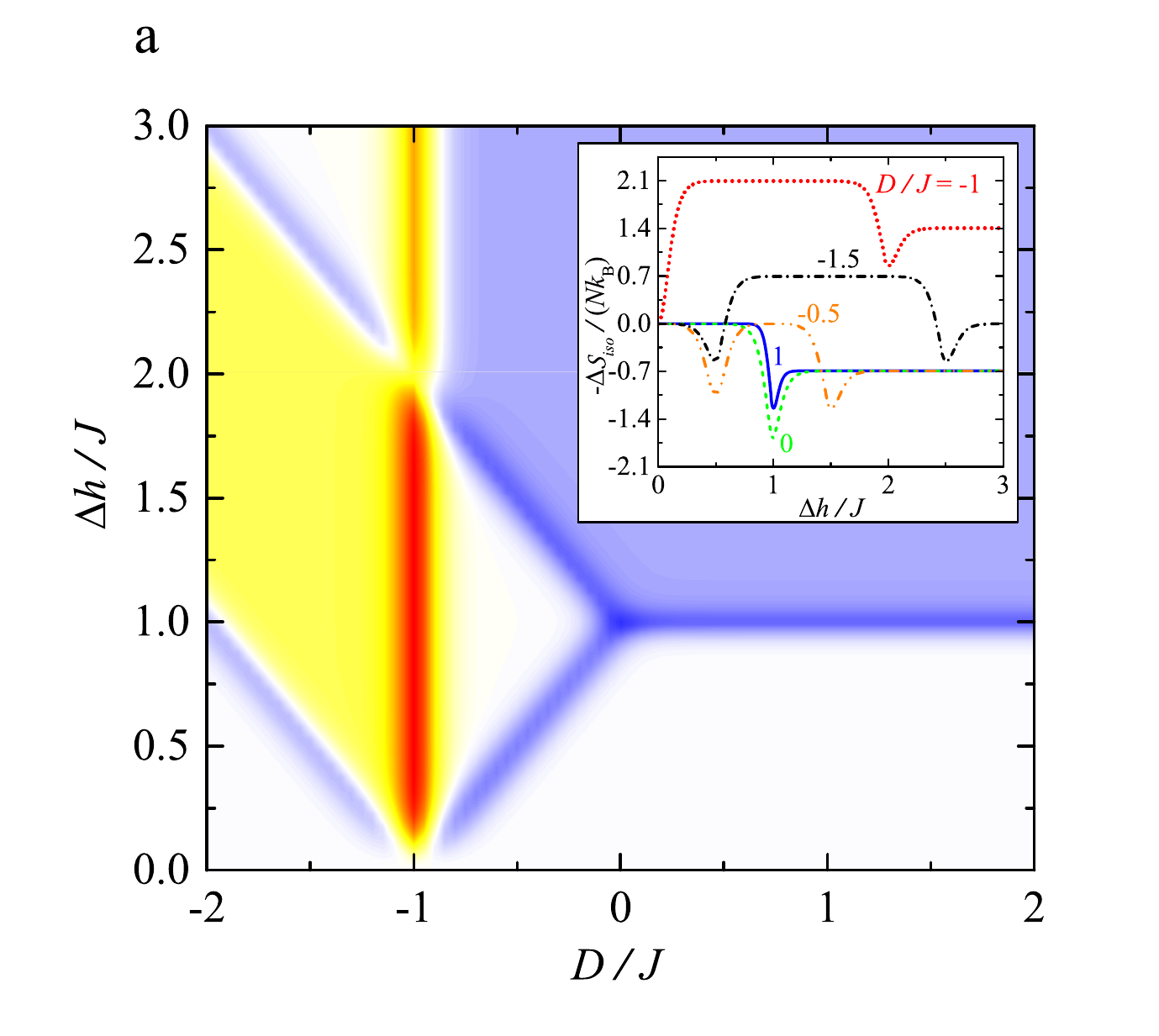}
  \hspace{-0.25cm}
  \includegraphics[width=0.46\textwidth]{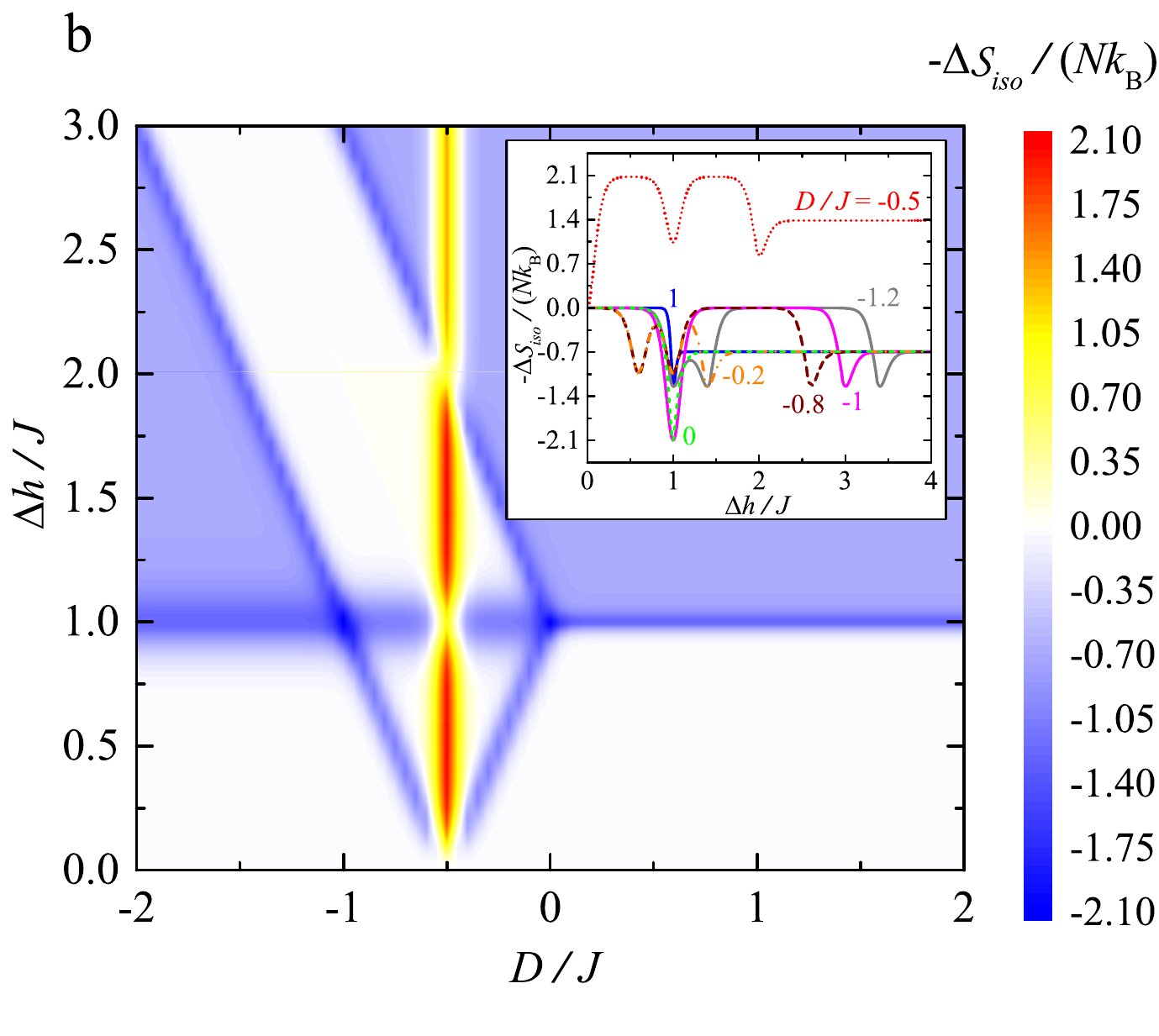}
\vspace{-0.25cm}
\caption{(Color online) Density plots of the isothermal magnetic entropy change $-\Delta{\cal S}_{iso}/(Nk_{\rm B})$ in the $D-\Delta h$ plane at the low enough temperature $k_{\rm B}T/J=0.05$ for the mixed spin-$(1/2,1)$ Ising model (figure~a) and the mixed spin-$(1/2,3/2)$ Ising model (figure~b). Insets show field dependencies of $-\Delta{\cal S}_{iso}/(Nk_{\rm B})$ for several representative values of the single-ion anisotropy parameter $D/J$.}
\label{fig:7}
\end{figure*}

An enhancement of the MCE in a vicinity of the first-order phase transitions during the isothermal magnetization process of the studied mixed-spin Ising model is obvious from Fig.~\ref{fig:7}. The low-temperature $-\Delta{\cal S}_{iso}(\Delta h)$ curves achieve pronounced local minima in a proximity of the field-induced phase transitions (see insets of Fig.~\ref{fig:7}), which indicates a rapid change in magnetocaloric properties of the model in these particular parameter regions. The observed minima become the sharper, the lower the temperature is. The magnitudes of these minima are given by a difference of the ground-state degeneracies (zero-temperature entropies) at the zero field and at individual field-induced phase transitions. 
Namely, if the frustrated P phase constitutes the zero-field ground state, the isothermal entropy change $-\Delta {\cal S}_{iso}\approx -0.548Nk_{\rm B}$ can be observed in a vicinity of the field-induced phase transitions. In accordance with the ground-state analysis, the aforementioned negative value of the isothermal entropy change indicating the inverse MCE can be found only in the mixed spin-$(1/2,1)$ Ising model for the single-ion anisotropy $D/J<-1$ (see the curve corresponding to $D/J=-1.5$ in inset of Fig.~\ref{fig:7}a). If LRO phases represents the respective zero-field ground state, more significant inverse MCE appears near the field-induced phase transitions upon the isothermal magnetization process. The observed phenomenon is proportional to the isothermal entropy changes $-\Delta {\cal S}_{iso}\approx -1.04Nk_{\rm B}$, $-1.241Nk_{\rm B}$, $-1.684Nk_{\rm B}$ and/or $-2.096Nk_{\rm B}$. The first (lowest) value $-\Delta {\cal S}_{iso}\approx -1.04Nk_{\rm B}$ can be identified in a vicinity of the first-order boundaries between different LRO phases (see e.g. the curves plotted for $D/J=-0.5$ in inset of Fig.~\ref{fig:7}a or $D/J=-0.8$, $-0.2$ in inset of Fig.~\ref{fig:7}b). The second isothermal entropy change $-\Delta {\cal S}_{iso}\approx -1.241Nk_{\rm B}$ corresponds to the LRO--P phase transitions (see the curve plotted for $D/J=1$ in inset of Fig.~\ref{fig:7}a and the curves corresponding to $D/J=-1.2$ and $1$ in inset of Fig.~\ref{fig:7}b). Finally, the last two values $-\Delta {\cal S}_{iso}\approx -1.684Nk_{\rm B}$ and $-2.096Nk_{\rm B}$ are associated with the triple point $[0, 1]$, which appears in the ground-state diagram of the mixed spin-$(1/2,1)$ model (see the curve plotted for $D/J=0$ in inset of Fig.~\ref{fig:7}a), and quadruple points $[-1, 1]$, $[0, 1]$, which may be observed in the zero-temperature phase diagram of the mixed spin-$(1/2,3/2)$ Ising model, respectively (see the curves depicted for $D/J=-1$ and $0$ in inset of Fig.~\ref{fig:7}b).
\begin{figure*}[t!]
\centering
  \includegraphics[width=0.46\textwidth]{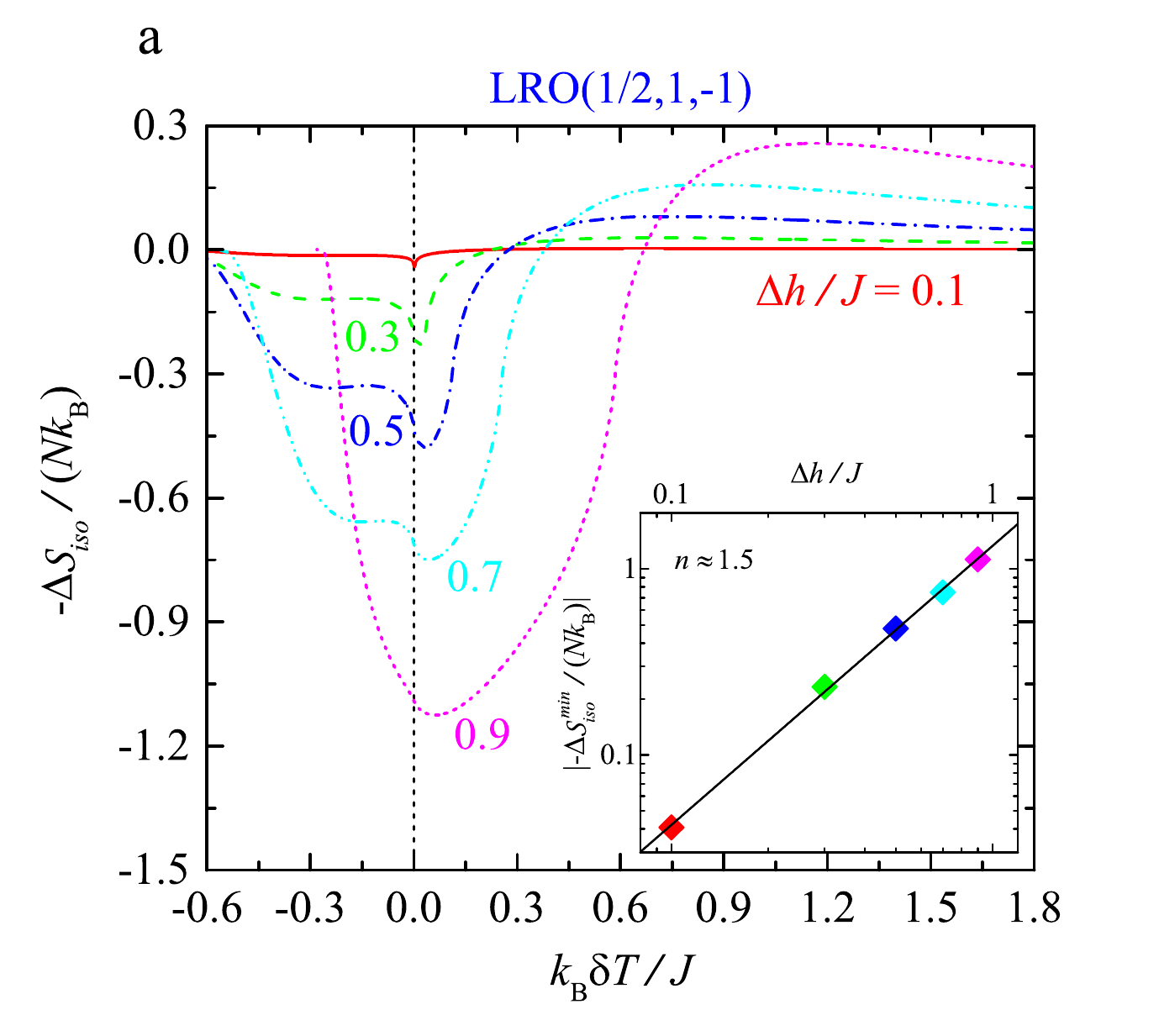}
    \hspace{-0.25cm}
  \includegraphics[width=0.46\textwidth]{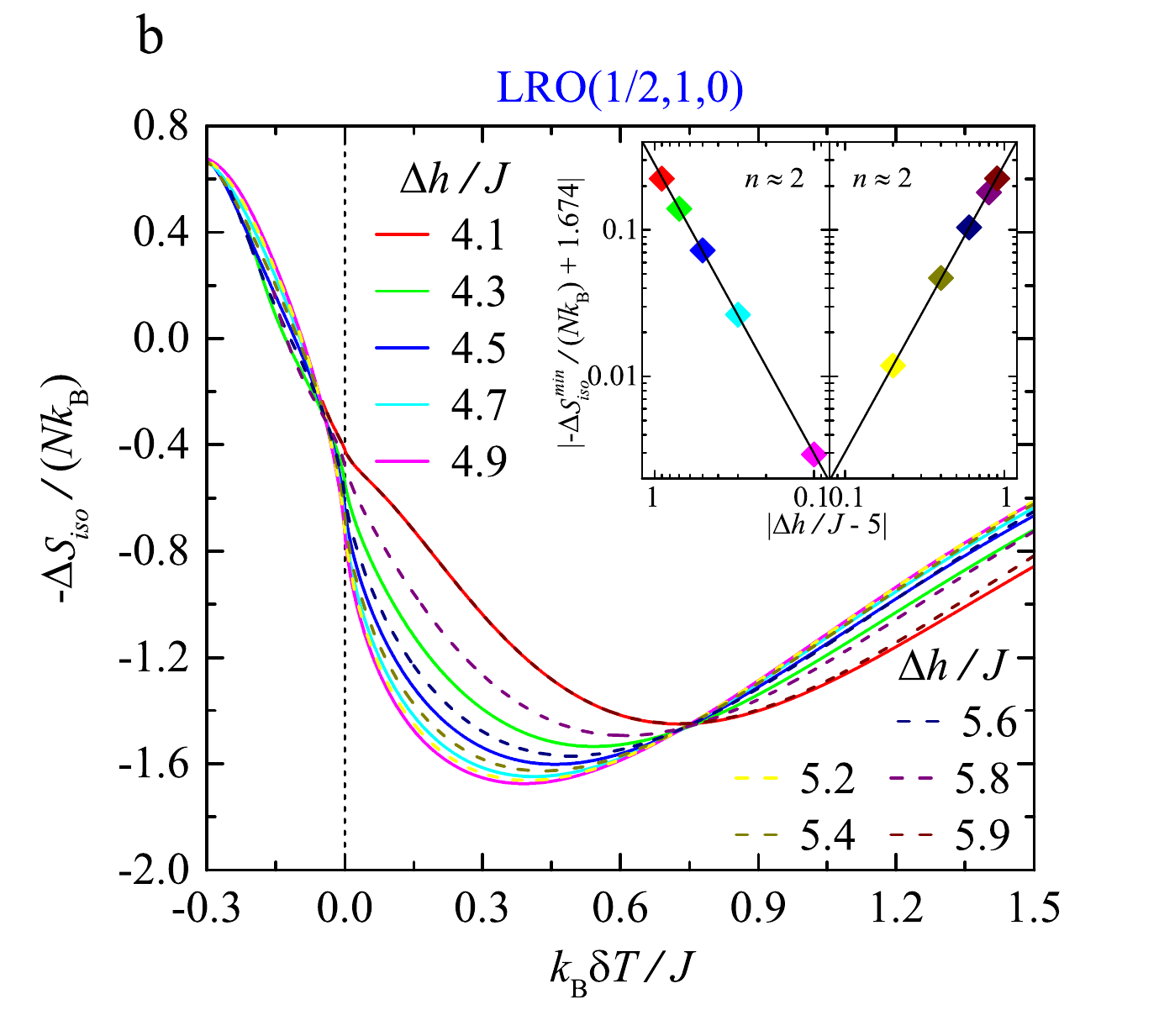}
\vspace{-0.25cm}
\caption{(Color online) Temperature dependencies of the isothermal magnetic entropy change $-\Delta{\cal S}_{iso}/(Nk_{\rm B})$ of the mixed spin-$(1/2,1)$ Ising model for various magnetic field changes $\Delta h\!:0\to h$ and the fixed single-ion anisotropy parameters $D/J = 5$ (figure~a) and $D/J = -5$ (figure~b). Insets show the field dependencies of the minimum entropy change $-\Delta{\cal S}_{iso}^{min}/(Nk_{\rm B})$ in log-log scale.}
\label{fig:8}
\end{figure*}

It is worth noticing that the non-zero isothermal entropy change can also be found in more
parameter space of the phase if its non-zero ground-state degeneracy differs from the zero-field one. As far as the mixed spin-$(1/2,1)$ Ising model is considered, there are two such parameter regions: the one occurs in the parameter range $D/J<-1$ within the phase LRO$(1/2,1,0)$, while another one is present in the range $D/J>-1$ within the disordered phase P$(\pm1/2,1,1)$ (compare Fig.~\ref{fig:7}a with Fig.~\ref{fig:2}a for clarity). The positive value of the isothermal entropy change $-\Delta {\cal S}_{iso} \approx 0.693Nk_{\rm B}$ detected in the former region unambiguously points to the moderate conventional MCE in the phase LRO$(1/2,1,0)$ if $D/J<-1$. Apparently, the observed phenome\-non is a result of the zero-field spin frustration of the $A$-sublatti\-ce in the phase P$(\pm1/2,0,0)$.
On the other hand, the disordered phase P$(\pm1/2,1,1)$ emerging in the region $D/J>-1$ is characterized by the opposite value of isothermal entropy change $-\Delta {\cal S}_{iso} \approx -0.693Nk_{\rm B}$, which  indicates the inverse MCE of moderate strength.
The same negative value of $-\Delta {\cal S}_{iso}$ can also be detected in both P phases, which occur in the ground-state phase diagram of the mixed spin-$(1/2,3/2)$ Ising model (compare Fig.~\ref{fig:7}b with Fig.~\ref{fig:2}b). Similarly as in the mixed spin-$(1/2,1)$ model, the observed phenomenon comes from the entropy increase due to spin frustration of the sublattice $A$. Last but not least, there are also two narrow regions with surprisingly large positive isothermal entropy changes $-\Delta {\cal S}_{iso} \approx 2.096Nk_{\rm B}$ and $2.079Nk_{\rm B}$. The higher value can be observed in the phase LRO$(1/2,1,0)$ for the special value of the single-ion anisotropy $D/J=-1$ (see Fig.~\ref{fig:7}a), while the lower one can be found in the phases LRO$(1/2,3/2,-1/2)$ and LRO$(1/2,3/2,1/2)$ if $D/J=-0.5$ (see Fig.~\ref{fig:7}b). Both aforementioned values of $-\Delta {\cal S}_{iso}$ indicate a relatively large conventional MCE in these particular regions due to a complete cancellation of the large zero-field degeneracies after turning on the magnetic field (see the ground-state analysis in the subsection~\ref{subsec:31}). The observed large conventional MCE is suppressed at critical fields corresponding to the field-induced phase transitions, but it does not completely disappear (see $-\Delta{\cal S}_{iso}(\Delta h)$ curves plotted for $D/J=-1$ and $-0.5$ in insets of Figs.~\ref{fig:7}a and~\ref{fig:7}b, respectively). In agreement with our expectation, the MCE again grows after crossing saturation fields due to spin frustration of the $A$-sublattice, but just to a~lower intensity $-\Delta {\cal S}_{iso} \approx 1.403Nk_{\rm B}$ and $1.381Nk_{\rm B}$. The former isothermal entropy change corresponds to the mixed spin-$(1/2,1)$ Ising model, while the latter one can be detected in the mixed spin-$(1/2,3/2)$ Ising model.

\begin{figure}[t!]
\vspace{0.25cm}
\centering
  \includegraphics[width=0.96\columnwidth]{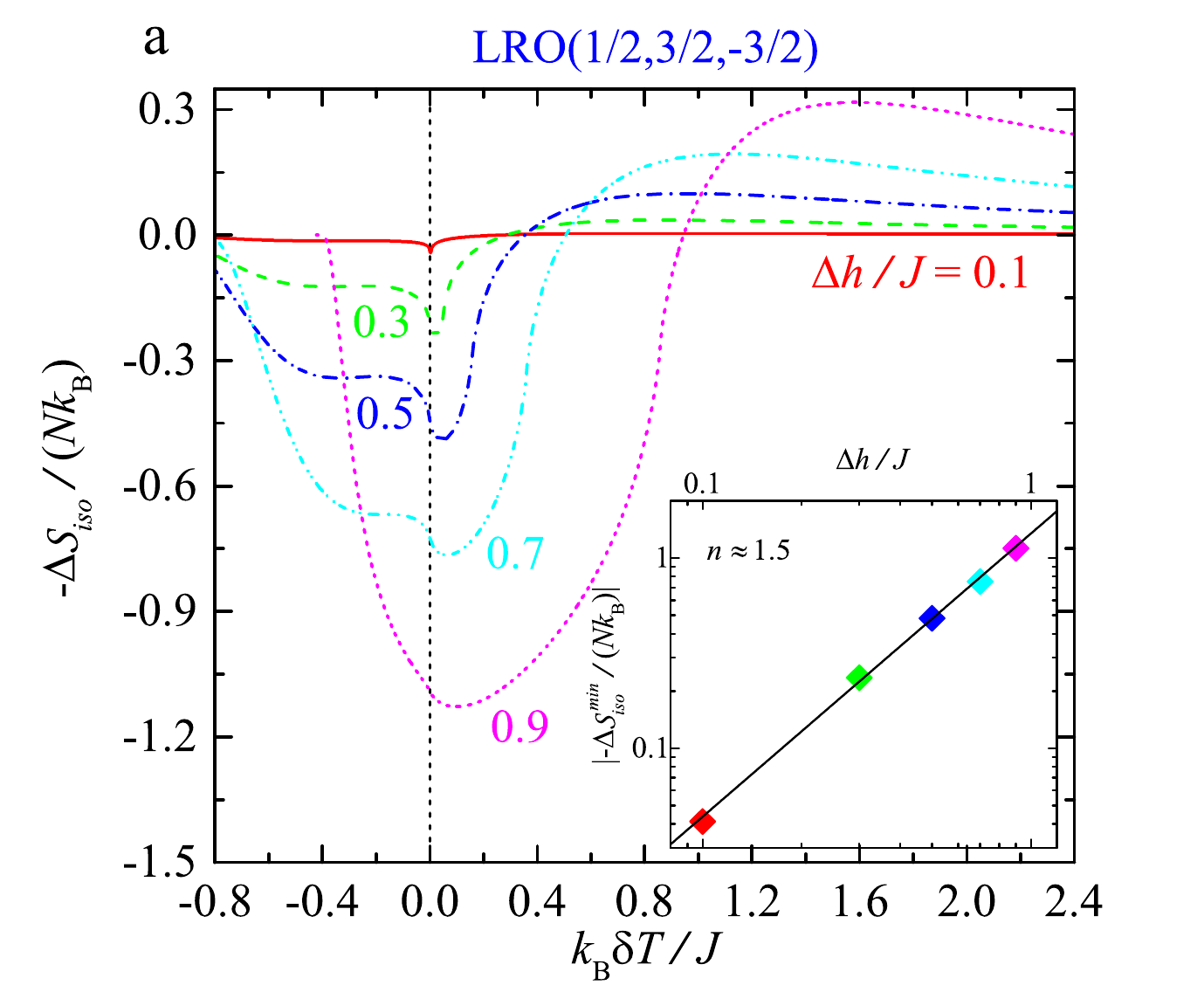}\vspace{1mm}
  \includegraphics[width=0.96\columnwidth]{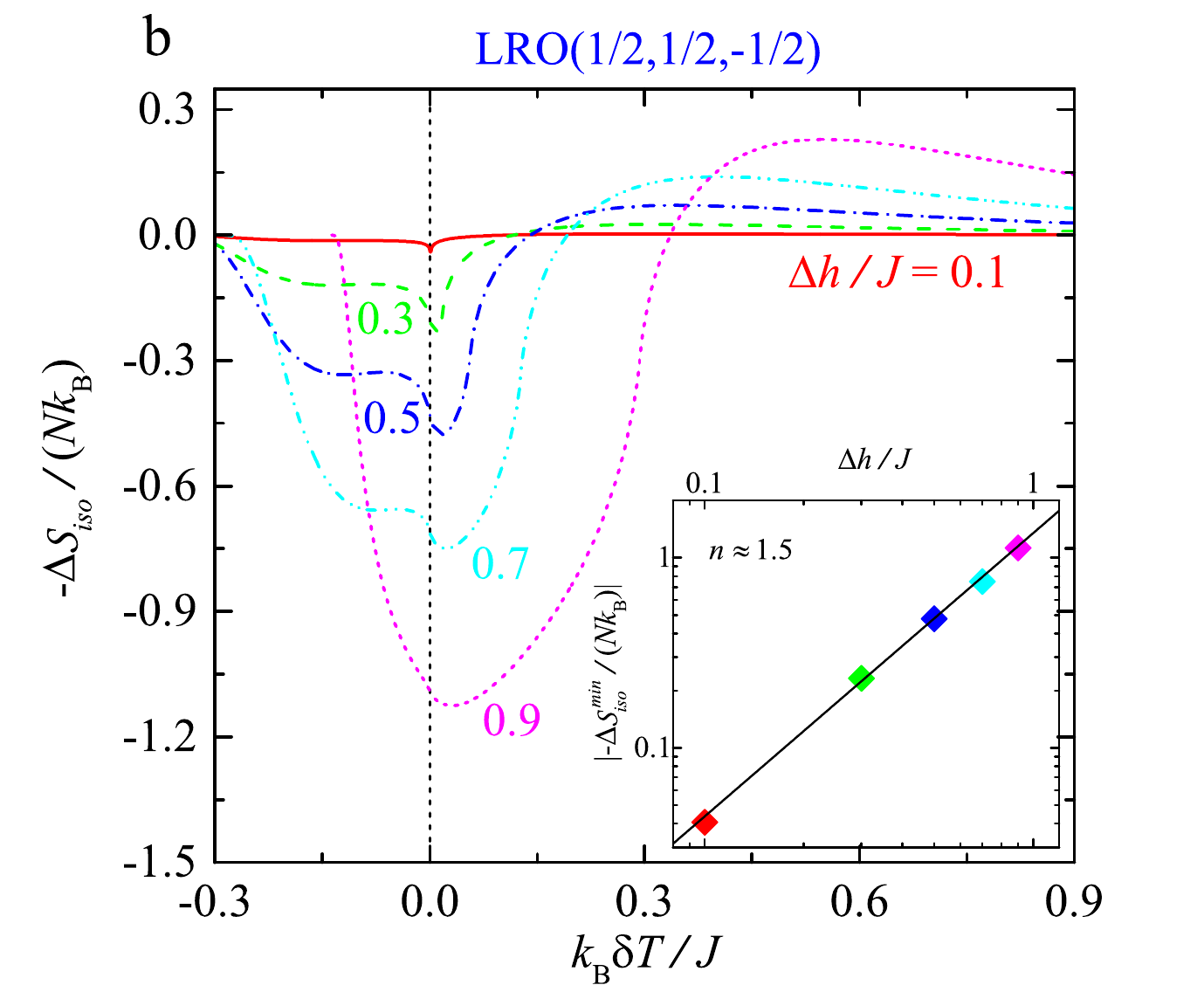}\vspace{1mm}
  \includegraphics[width=0.96\columnwidth]{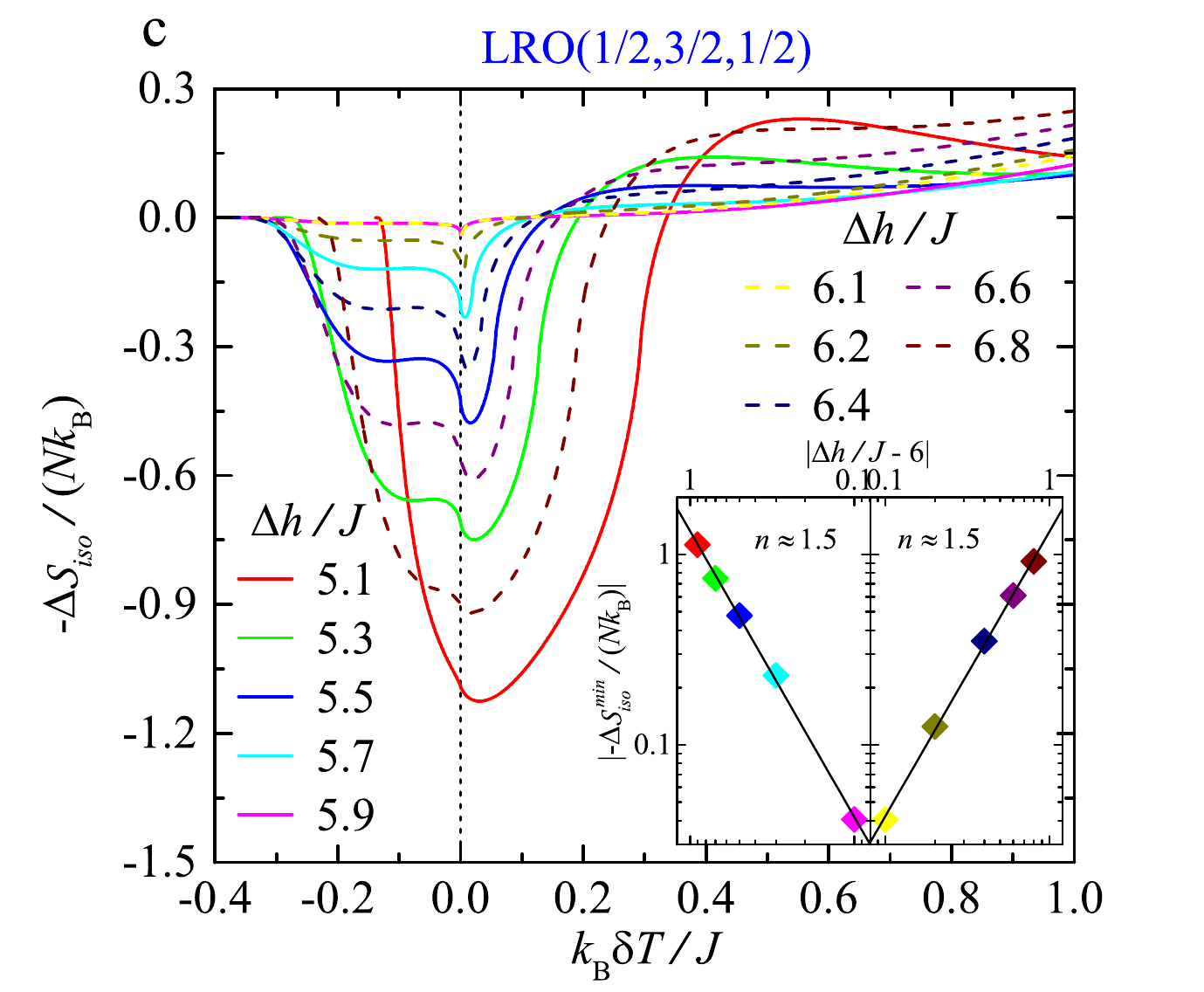}
\vspace{-0.15cm}
\caption{(Color online) Temperature dependencies of the isothermal magnetic entropy change $-\Delta{\cal S}_{iso}/(Nk_{\rm B})$ of the mixed spin-$(1/2,3/2)$ Ising model for various magnetic field changes $\Delta h\!:0\to h$ and the fixed single-ion anisotropy parameters $D/J = 3$ (figure~a) and $D/J = -3$ (figures~b, c). Insets show the field dependencies of the minimum entropy change $-\Delta{\cal S}_{iso}^{min}/(Nk_{\rm B})$ in log-log scale.}
\label{fig:9}
\vspace{-0.75cm}
\end{figure}

Figures~\ref{fig:8} and~\ref{fig:9} display temperature dependencies of the isothermal magnetic entropy change for several magnetic-field changes $\Delta h\!\!:0\to h$, which clarify magnetocaloric properties of the mixed-spin Ising model in a vicinity of the continuous, temperature-induced phase transitions. Since the critical temperature of the studied model is significantly varied with the changing external magnetic field, temperature axes of the figures are rescaled to $k_{\rm B}\delta T/J = k_{\rm B}(T - T_c)/J$.
It is apparent from Figs.~\ref{fig:8} and~\ref{fig:9} that the temperature dependencies of the isothermal entropy change $-\Delta {\cal S}_{iso}$ exhibit negative minima at the temperatures $T\gtrsim T_c$, which clearly points to the inverse MCE slightly above temperature-induced phase transitions. In general, the observed inverse MCE has an universal field evolution following the power law $|-\Delta {\cal S}_{iso}^{min}|\propto h^n$~\cite{Fra06, Fra09a, Fra09b, Fra10}. The value of the local exponent $n$ can be easily determined from slopes of the linear functions fitted to the appropriate $-\Delta {\cal S}_{iso}^{min}$ data, which are depicted in insets of Figs.~\ref{fig:8} and~\ref{fig:9} in log-log scale. Interestingly,  $n$ is quite sensitive on the ground-state arrangement of decorating spins. To be specific, if decorating spins occupy solely magnetic states at the absolute zero temperature, then the exponent takes the value $n\approx 1.5$ regardless of the magnitude of decorating spins, magnetic-field change, and the single-ion anisotropy (see insets of Fig.~\ref{fig:8}a and Fig.~\ref{fig:9}). On the other hand, the local exponent takes the value $n\approx 2$ if only a half of all decorating spins in the model occupies magnetic states and other half is in a non-magnetic state (see inset of Fig.~\ref{fig:8}b). Again, the observed value of $n$ does not depend neither on the magnitude of integer decorating spins, nor the parameters $D$ and $\Delta h$.

\section{Conclusions}
\label{sec:4}

The present paper deals with magnetic and magnetocaloric properties of the mixed-spin Ising model on a decorated triangular lattice in a longitudinal magnetic field, which is exactly solvable by performing the generalized decoration-iteration \linebreak transformation~\cite{Fis59,Dom60,Syo72,Str10} when some specific restrictions on mo\-del parameters are assumed.

The numerical results obtained for two particular values of the decorating spins $s = 1$ and $s = 3/2$ demonstrate that a mutual interplay between model parameters and the applied magnetic field gives rise to various long-range ordered and disordered paramagnetic ground states. Macroscopic degeneracies of the latter group of phases, phase boundaries and their crossing points have been discussed. The highest macroscopic degeneracy has been found at those coexistence points, where up to seven different types of spin configurations are in thermodynamic equilibrium. The zero-temperature phase diagram of the mixed spin-$(1/2,1)$ Ising model contains one such point, while the zero-temperature phase diagram of the mixed spin-$(1/2,3/2)$ Ising model contains two such points. The observed diversity of the ground-state spin arrangements of both investigated models reflects itself in a rich variety of magnetization process involving several magnetization plateaux, as well as in interesting critical behaviour of these systems.

The investigation of the isothermal entropy change points to the existence of the inverse and conventional MCE in the vicinity of the discontinuous field-induced phase transitions and their crossing points, as well as near the critical temperature. At low enough temperatures, the inverse MCE is the most pronounced near crossing points of the first-order phase transitions, while the large conventional MCE arises as soon as the zero-field degeneracy is lifted by the magnetic field. The inverse MCE  observed slightly above critical temperature of the model follows the well-known power law $|-\Delta {\cal S}_{iso}^{min}|\propto h^n$, where $n$ is quite sensitive to the spin ordering in the ground state.

\section*{Acknowledgments}
This work was financially supported by Ministry of Education, Science, Research and
Sport of the Slovak Republic under the grant VEGA 1/0043/16, by the Slovak
Research and Development Agency under the Contract No. APVV-16-0186, and by Faculty of Mechanical Engineering of Technical University of Ko\v{s}ice under the internal grant of young scientists.

\end{document}